# Spectroscopic Evidence for the Direct Involvement of Local Moments in the Pairing Process of the Heavy-Fermion Superconductor CeCoIn$_5$

K. Shrestha[1,†], S. Zhang[1,2], L. H. Greene[1,2], Y. Lai[1,2], R. E. Baumbach[1,2], K. Sasmal[3], M. B. Maple[3], and W. K. Park[1,*]

[1]*National High Magnetic Field Laboratory, Florida State University, Florida, 32310, USA*
[2]*Department of Physics, Florida State University, Florida, 32306, USA*
[3]*Department of Physics, University of California, San Diego, California, 92093, USA*

Abstract

The microscopic mechanism for electron pairing in heavy-fermion superconductors remains a major challenge in quantum materials. Some form of magnetic mediation is widely accepted with spin fluctuations as a prime candidate. A novel mechanism, "composite pairing" based on the cooperative two-channel Kondo effect directly involving the *f*-electron moments has also been proposed for some heavy fermion compounds including CeCoIn$_5$. The origin of the spin resonance peak observed in neutron scattering measurements on CeCoIn$_5$ is still controversial and the corresponding hump-dip structure in the tunneling conductance is missing. This is in contrast to the cuprate and Fe-based high-temperature superconductors, where both characteristic signatures are observed, indicating spin fluctuations are likely involved in the pairing process. Here, we report results from planar tunneling spectroscopy along three major crystallographic orientations of CeCoIn$_5$ over wide ranges of temperature and magnetic field. The pairing gap opens at $T_p \sim 5$ K, well above the bulk $T_c$ = 2.3 K, and its directional dependence is consistent with $d_{x^2-y^2}$ symmetry. With increasing magnetic field, this pairing gap is suppressed as expected but, intriguingly, a gaplike structure emerges smoothly, increasing linearly up to the highest field applied. This field-induced gaplike feature is only observed below $T_p$. The concomitant appearance of the pairing gap and the field-induced gaplike feature, along with its linear increase with field, indicates that the *f*-electron local moments are directly involved in the pairing process in CeCoIn$_5$.

[†]Present address: Department of Chemistry and Physics, West Texas A&M University, 2501 4th Ave, Canyon, Texas 79016, USA.

[*]To whom correspondence should be addressed. Email: wkpark@magnet.fsu.edu.

# I. INTRODUCTION

The unconventional heavy-fermion superconductor CeCoIn$_5$ has the critical temperature ($T_c$) of 2.3 K and a $d_{x^2-y^2}$ superconducting (SC) order parameter symmetry [1,2]. Since its discovery, the glue for electron pairing has been suggested to entail magnetic mediation, but the precise mechanism remains to be revealed. In the cuprate and Fe-based high-temperature superconductors (HTS) [3], antiferromagnetic spin fluctuations [4-6] have been identified as playing a major role in the Cooper pairing based on the observation of the neutron spin resonance and the corresponding signature in tunneling conductance [7-11], which is how the phonon-mediated pairing mechanism was confirmed in conventional superconductors [12-15]. This is not the case for CeCoIn$_5$ as the origin of the neutron resonance peak at $\Omega_{res}$ = 0.6 meV remains controversial [16-18], and there is no corresponding feature in the tunneling data [19-23], suggesting a different pairing interaction. The anomalous magnetic susceptibility in CeCoIn$_5$, which indicates the presence of un(der)-screened moments [24] down to $T_c$, led Coleman and co-workers to propose a new pairing mechanism [25,26] in which "composite" pairs are formed between local moments and conduction electrons through a cooperative two-channel Kondo effect [27-29].

Here we report results from planar tunneling spectroscopy (PTS) [30] measurements on CeCoIn$_5$. In addition to the main SC phase with $d_{x^2-y^2}$ symmetry in CeCoIn$_5$, there exists another distinct phase, the Q-phase (previously thought to be the Fulde-Ferrell-Larkin-Ovchinnikov phase [31-33]), appearing only in a limited region of the phase diagram (low temperature and high magnetic field) [34-36]. In this paper, we focus our discussion on the overall temperature and field dependences since no noticeable changes are observed in the Q-phase. Our detailed and reproducible tunneling conductance spectra provide strong evidence for: i) the existence of preformed pairs well above $T_c$; and ii) the direct involvement of localized *f*-electron moments in the pairing process. Surprisingly, local physics manifested via Kondo resonance appears to play a key role in the superconductivity in this compound.

# II. EXPERIMENTS

The CeCoIn$_5$ single crystals used in our studies were grown by three independent groups using the flux method [37]. High-quality crystals, based on both magnetization and resistivity measurements, were chosen and cut to have the surface orientation along three major crystallographic axes, namely, [001], [100], and [110], as determined by single crystal x-ray diffraction [38]. They were fixed on epoxy (Stycast® 2850-FT) molds and then polished down to 1.0 – 1.5 nm peak-to-dip smoothness (Fig. S1 in [39]). The superconductor/insulator/superconductor (S-I-S') tunnel junctions were prepared by depositing a 2.0 – 2.5 nm-thick aluminum layer on the polished crystal surface, followed by subsequent plasma oxidation, then deposition of lead (Pb) strips as counter electrodes (Fig. S2 in [39]).

Measurements of the differential tunneling conductance across the junction, $G(V) \equiv \frac{dI}{dV}$, were carried out using the four-probe lock-in technique over wide ranges of temperature ($T$, down to 20 mK) and magnetic field ($H$, up to 18 T). Here, V is 'Sample Bias' voltage applied to the CeCoIn$_5$. Only the conductance spectra from high-quality junctions, determined by the sharpness of the Pb coherence peaks and phonon features (Figs. S3 – S5 in [39]), are reported here. The conductance spectra of CeCoIn$_5$, as presented in the main text, were obtained by driving the Pb normal with a small magnetic field (H = 0.2 T). Unless otherwise specified throughout this paper, magnetic fields were applied perpendicular to the junction plane. We define the normalized conductance in two different ways: i) $G_n(V) \equiv G(V)/G(-V_{max})$, where $-V_{max}$ is the negative maximum bias voltage; ii) $G_b(V) \equiv G(V)/G_{bg}(V)$, where the background conductance $G_{bg}(V)$ is obtained from a polynomial fitting of the G(V) in the high-bias region. The typical junction resistance was $R_J$ = (20 – 50) Ω, and the product $R_J·A$ = was 10-20 Ωmm$^2$, where A is the junction area. See Supplemental Material ([39] Sect. 1. Materials and Methods) for additional details.

### III. SUPERCONDUCTING ORDER PARAMETER AND PRE-FORMED PAIRS

The tunneling conductance data taken at 20 mK are plotted in Fig. 1. While both (001) and (100) junctions show sharp coherence peaks, the (110) junction exhibits a pronounced zero-bias conductance peak (ZBCP). To extract the SC gap, Δ, the $G_b(V)$ curves from the (001) and (100) junctions are analyzed by fitting to the *d*-wave Blonder-Tinkham-Klapwijk (BTK) model [40,41] with three adjustable parameters: Δ, Γ, and Z (Sect. 2 in [39]). Here, Γ is the quasiparticle lifetime broadening parameter and Z represents the dimensionless barrier strength. Unlike the (100) junction in Fig. 1(b), the U-shape subgap-conductance of the (001) junction can't be replicated by the d-wave BTK model as seen in Fig. 1(a), possibly due to the tunneling cone effect (Fig. S6 in [39]). The extracted Δ value is 0.66 meV and 0.54 meV for the [001] and [100] directions, respectively, falling in the range reported in the literature [1,19-23]. The ZBCP seen in the (110) junction can be a characteristic feature of a nodal junction on a *d*-wave superconductor, arising from surface bound states formed due to the sign-changing nature of the *d*-wave order parameter, known as Andreev bound states (ABS) [42-45]. Thus, overall, the anisotropy in our tunneling conductance agrees with the well-established $d_{x2-y2}$–wave pairing symmetry in CeCoIn$_5$ [1,2]. On a closer look, the ZBCP consists of two structures: A wider peak of Lorentzian shape as shown by the red solid line and a narrower peak sitting on top of the former. The wider peak itself is not due to ABS, as discussed later regarding its magnetic field dependence. The narrower structure can be seen more clearly in the left inset of Fig. 1(c), plotting the $G_b(V)$ further normalized by the Lorentzian background [46]. It consists of slightly split peaks, reminiscent of the Doppler shift of ABS under a magnetic field (0.2 T) [47,48]. The gap edge expected to be seen along with the ZBCP is not apparent presumably because the peak isn't narrow enough compared to the small Δ in CeCoIn$_5$. The detailed behavior of this possibly ABS-originated ZBCP remains to be further investigated.

Figure 2 shows temperature evolution of the tunneling conductance along the three directions in both waterfall plots, (a) – (c), and color contour maps, (d) – (f). For a (001) junction in Fig. 2(a), with decreasing temperature, a broad ZBCP emerges and gradually grows until $T_p$ ~ 5 K, where it begins to split, as shown more clearly in the left inset, smoothly evolving into a SC gap that turns into well-defined coherence peaks at low temperature. Thus, we interpret the splitting of the ZBCP as due to the opening of a gap in the single particle spectrum caused by the pairing of conduction electrons. The right inset displays the ZBC vs. $T$, which shows a sharp drop at $T$ ~ 5 K, further confirming the temperature scale, $T_p$ ~ 5 K. The pairing gap persisting above $T_c$ can also be seen in the color-contour map in Fig. 2(d). This is in agreement with previous scanning tunneling spectroscopy (STS) [19,20], resistivity [49], and thermal conductivity [50] studies that identified a pseudogap in this temperature range. Four-fold oscillations in the field-angle dependent thermal conductivity [50], an evidence for the $d_{x2-y2}$ pairing symmetry in CeCoIn$_5$, were observed up to $T$ = 3.2 K, implying that the pairing gap above $T_c$ has the same symmetry as that below $T_c$. The persistence of the Curie-Weiss temperature dependence of the DC magnetic susceptibility, $\chi$, another bulk property, down to just above $T_c$ was one of the key experimental observations underlying the theoretical proposal for a novel pairing mechanism in CeCoIn$_5$ [25,26]. On a close look, we notice that $\chi$ exhibits a slight but clear slope decrease below ~$T_p$ in its Curie-Weiss plot vs. $1/T$ (Fig. S7 in [39]). As $\chi$ in CeCoIn$_5$ is primarily due to the Ce$^{3+}$ ions or localized $4f^1$ electrons, this slope decrease concomitant with the opening of the pairing gap as seen in our tunneling spectroscopy suggests that, indeed, the Ce-$4f$ electrons might be directly involved in the pairing process [25,26]. The pairing gap above $T_c$ in CeCoIn$_5$ is reminiscent of the pseudogap in high-$T_c$ cuprates [51,52]. And, the continuous evolution of the pairing gap feature crossing the $T_c$ implies that preformed pairs exist in the pseudogap region below $T_p$, whose nature is further discussed later.

In the above discussion, we have shown that the onset of the pairing gap at $T_p > T_c$ evidenced in our single electron tunneling spectra is also consistent with other bulk properties including resistivity, thermal conductivity, and magnetic susceptibility. On the other hand, there is no such evidence in the specific heat ($C$) [24] or Andreev reflection (AR) measurements [1]. This can be understood as follows. Since $C = -T\frac{\partial^2 F}{\partial T^2}$, where $F$ is the free energy, it is directly tied to the SC order parameter, $\Psi_{SC} = |\Psi|e^{i\varphi}$, where $\varphi$ is the phase factor. Preformed pairs in the pseudogap region are not yet condensed into the same ground state, so $\Psi_{SC} = 0$, hence there would be no signature in $C$ across $T_p$. In the case of AR, if an electron of energy $E$ from the normal metal is injected into the superconductor, the phase change during its reflection as a hole is given by $\Phi = \varphi + \cos^{-1}(E/\Delta)$. It is generally believed that the AR conductance is detectable because the superconductor has a well-defined order parameter with a definite $\varphi$. Thus, the reason why the AR conductance is zero in the preformed pair state of CeCoIn$_5$ [1] could be because $\varphi$ is random among the pairs, that is, they remain incoherent down to $T_c$. It is an open question whether AR can still occur off individual pairs [53] but incoherently, resulting in overall cancellation in typical time-averaged measurements such as differential conductance or it can't occur at all until full phase coherence is reached below $T_c$.

Figures 2(b) & 2(e) show the temperature evolution of $G_b(V)$ for a (100) junction. The pairing gap feature persists above $T_c$ albeit weaker, similarly to that of the (001) junction in Fig. 2(a). This is also evidenced by the drop of the ZBC below ~ 4 K, as shown in the inset of Fig. 2(b). It is notable that the pairing gap feature in this junction emerges out of a zero-bias conductance dip (ZBCD), in contrast to the gap emerging out of a ZBCP in the (001) junction (Fig. 2(a)). Empirically, (001) junctions have been observed to show a ZBCP more frequently than a ZBCD, whereas it is opposite for (100) junctions. (110) junctions always exhibit a ZBCP. While further investigations are necessary to pin down the exact origin for these discrepant behaviors, here we discuss some clues. In a Kondo system (whether single impurity or lattice), electrons can co-tunnel into the conduction band and the localized state (orbital) [54-57], resulting in a Fano resonance with the conductance shape strongly depending on the Fano parameter, $q_F$ [54,58]. Thus, the variation of the conductance shape can be attributed to the $q_F$ value: A peak (dip) for large (small) $q_F$ due to the predominant tunneling into the localized orbital (conduction band) in these junctions (Fig. S8 in [39]). Related to this, we note the ZBC has a finite value even at very low temperatures in both the (001) and (100) junctions, as shown in Fig. 1. Our smallest observed ZBC is 19% (not shown) of the high-bias conductance, substantially smaller than that (~50%) reported in most of the previous STS measurements [19,21-23]. A finite conductance within the SC gap at such a low temperature cannot be explained by the thermal population effect alone. Based on our observation of both ZBCP and ZBCD as mentioned above (e.g., see Fig. S9 in [39]), we speculate that it may be associated with the existence of non-trivial tunneling channels, an intrinsic property of CeCoIn$_5$, as detailed below. The temperature evolution of a (110) junction is shown in Figs. 2(c) & 2(f) from 0.4 K up to 30 K (see Fig. S10 in [39] for another set of conductance spectra). The ZBCP becomes wider with increasing temperature with the ZBC showing a logarithmic dependence in the intermediate temperature range, reminiscent of a Kondo resonance. The persistent observation of a ZBCP in all (110) junctions suggests that the Kondo resonant tunneling off the localized Ce $4f^1$ moments is enhanced in this direction compared to other directions. This is in agreement with a recent report [59] that the lobe direction of the ground state $4f$ orbital in CeCoIn$_5$ is [110]. The ZBCPs observed in some non-nodal junctions (e.g., Fig. S11 in [39]) may have a similar origin, presumably caused by the crystal surface' atomic-scale structure being favorable for a Kondo resonant tunneling, i.e., along the $4f$ orbital's lobe direction. Within this local picture, the ZBCDs observed in the other non-nodal

junctions can also be understood as due to a dominant tunneling along the 4*f* orbital's nodal direction, namely, into the conduction band, resulting in a ZBCD due to an anti-resonance.

To determine the temperature dependence of $\Delta$, we have analyzed the conductance data displayed in Fig. 3(a), which were taken from a (001) junction. For simplicity, $G_b(V)$ is obtained by dividing out $G(V)$ at each temperature with $G(V)$ at 5 K that shows a ZBCD, which is then fit to the *d*-wave BTK model [40,41]. Best fits are obtained with Z kept to a constant value of 5.0, well in the tunneling limit, and plotted in Fig. 3(b). The temperature dependence of extracted $\Delta$ and $\Gamma$ is shown in Fig. 3(c). At $T = 0.4$ K, $\Delta = 0.87$ meV, again falling in the range reported in the literature [1,19-23]. Note $\Delta$ decreases gradually with *T*, has a finite value above $T_c$, and tends to zero only at $T \sim 5$ K (dashed line). Meanwhile, $\Gamma$ increases with *T*, as expected. It is notable that at $T_c$, $\Delta \sim 3\Gamma$ within the error bar. A similar scaling behavior between $\Delta$ and $\Gamma$ has been reported in photoemission studies of some high-$T_c$ cuprates [60] and can also be seen in an STS study of CeCoIn$_5$ [20]. Assuming that $\Gamma$ is related to a pair-breaking scattering rate, $\Gamma = \hbar/\tau$, where $\tau$ is the lifetime of the Cooper pair, this result can be interpreted as follows: with decreasing temperature below $T_p$, the density of Cooper pairs increases until a critical density is reached and condensation occurs [60]. Thus, the relationship $3\Gamma(T)/\Delta(T)|_{T=T_c} = 1$ appears to define $T_c$. However, it should be noted that not all junctions show exactly the same scaling behavior as in this junction. A more in-depth analysis is required to address whether this is due to the tunneling spectrum being affected by the Kondo (anti-)resonance, as discussed above.

**IV. ANOMALOUS EVOLUTION OF THE PAIRING GAP UNDER MAGNETIC FIELD**

The magnetic field evolution of the tunneling conductance is shown in Fig. 4 for all three directions. For the (001) and (100) junctions, the application of an external magnetic field suppresses the pairing gap feature gradually, as expected, but an intriguing field-induced gaplike feature (FIG) emerges at higher fields. We stress that the FIG appears even before the closing of the pairing gap at the upper critical field ($H_{c2} = 4.95$ T and 11.8 T along the [001] and [100] directions, respectively) [31]. Note that both the depth and width of the FIG increase with increasing field up to 18 T, the highest field applied. Note also that the FIG is observed below $T_c$ (Figs. 4(a)-(b)) and above $T_c$ (Figs. 4(d)-(e)). In both orientations, the tunneling conductance shows a crossover from the SC gap feature to the FIG. In contrast, the nodal junction exhibits a ZBCP both below and above $T_c$ with no apparent pairing gap, as already seen in Figs. 1 & 2. At $T = 20$ mK, the top part of the ZBCP is split at $H = 0.2$ T, which could be a signature for the Doppler shift of ABS as discussed in Fig. 1(c). However, the major part of the ZBCP can't be a signature for ABS since it splits persistently all the way up to 18 T, well above $H_{c2}$. Instead, it may originate from a Kondo resonance, as mentioned earlier, a part of the hybridization process leading to the lattice coherence. Indeed, the sharp

ZBCP at low temperature is seen to grow out of a broad ZBCP that begins to appear below 45 K (see Fig. S10 in [39]), widely known as the coherence temperature in CeCoIn$_5$ [24]. At $T = 5$ K, the splitting is not observed until $H \approx 14$ T, whose exact understanding beyond the thermal population effect requires further investigations since the splitting must be intimately tied to the exact origin of the ZBCP.

Prior to conducting a quantitative analysis of the field dependence just described above, it is important to determine whether the FIG is due to an extrinsic or intrinsic effect, and if intrinsic, whether it reflects the surface or bulk property. Based on the data shown in Fig. 4, three possibilities can be considered (see Fig. S2(c) in [39]): case A – extrinsic magnetic moments in the barrier or at the interface; case B – surface Ce$^{3+}$ ions acting as Kondo impurities; case C – bulk effect. A magnetic moment in the tunnel barrier or at the interface can cause a (Kondo) resonant tunneling at the Fermi level, showing up as a ZBCP, and an applied magnetic field causes a Zeeman splitting, as observed frequently in PTS and STS [61-63] and explained by the Anderson-Appelbaum (AA) theory [64,65]. The ZBCP observed in the nodal junction on CeCoIn$_5$ and its splitting under an applied magnetic field is reminiscent of this single impurity Kondo effect. However, such a ZBCP has never been observed in our junctions prepared on many other materials than CeCoIn$_5$ using the same procedure to form AlO$_x$ [66] and is extremely rarely reported in the literature [61], albeit the possibility of forming magnetic moments in AlO$_x$ [67]. The FIG in CeCoIn$_5$ has also been reported in recent STS studies [21,23], in which tunneling conductance was measured on a surface cleaved freshly in vacuum, so such magnetic moments of extrinsic origin can be ruled out. Thus, we are left with the other two possibilities for the intrinsic origin of the FIG. For case B, a metallic point-contact junction on CeCoIn$_5$ is expected to exhibit a ZBCD but such a signature due to single impurity Kondo scattering [68] has never been observed in our measurements on all three surfaces of CeCoIn$_5$ [1,69]. In addition, our analysis of the ZBCP using the Frota function [46] and its temperature evolution in terms of interaction-induced broadening within the strong coupling regime [63,70] (see Fig. S13 in [39]) suggests that case C is more likely than case B. Thus, we conclude the FIG reflects a bulk property of CeCoIn$_5$.

For further analysis of the FIG, in Figs. 4(g) – (i), we plot the low-temperature field evolution of the nominal peak position, $V_p$, corresponding to the SC gap at low fields and the FIG at high fields. For the (001) and (100) junctions, $V_p$ decreases gradually as expected for a pairing gap, but only up to the crossover field, $H_{cr} \approx 4.0$ T, above which it increases linearly due to the FIG's takeover. This crossover behavior is also seen in the field dependence of the ZBC (Fig. S12 in [39]). It is interesting that, although the FIG is dominant above $H_{cr}$, the pairing gap along [100] is seen to persist up to $H_{c2}$ (Fig. S14 in [39]), as shown by $V_{ps}$ in Fig. 4(h) at which the conductance slope change is clearly observed. If the $V_p$ below $H_{cr}$ is extrapolated to the field axis using the field dependence of the pairing gap given by the Ginzburg-Landau (GL) theory [71], $\Delta(H) = \Delta(0)\sqrt{1 - (H/H_{c2})^2}$, where $\Delta(0)$ is the gap at zero field, $V_p = 0$ at $H \approx H_{c2}$ in both directions. This confirms that the pairing gap observed in our PTS represents a bulk property. Unlike the (001) and (100) junctions, $V_p$ in the nodal junction increases linearly up to the highest field applied.

The nominal peak position ($V_p$) increases linearly above $H_{cr}$ in the non-nodal and at all fields in the nodal junctions and, extrapolating from high field, ($H$, $V_p$) approaches (0, 0) (green dash-dotted lines). The (001) junction shows a slight offset when extrapolated to zero field. This may be explained by the large smearing effect ($\Gamma$) in this junction, which can be inferred from the larger $\Gamma/\Delta$ ratio at zero field compared to the (100) junction (see the Fig. 4 caption). The linear increase of the field-induced splitting is reminiscent of the Zeeman effect: $eV_p = E_Z = \frac{1}{2} g\mu_B H$, where $E_Z$ is the Zeeman energy, g is the Landé g-factor, and $\mu_B$ is the Bohr magneton. However, it is well known that, for tunneling into single Kondo impurities [61-63], the slope of a $V_p$ vs. $H$ plot gives a wrong g-value. Instead, the g-factor can be determined reasonably accurately by taking the point where the slope of the conductance is largest, $V_s$. This is also justified from our simulation (see Fig. S15(a) in [39]), so we determine $V_s$ as a function of the field (Fig. S15(b) in [39]). The $V_s$ values are plotted in Figs. 4(g) – 4(i), in which ($H$, $V_s$) extrapolates to (0, 0) for all three junctions (see black dashed lines). Since $V_s$ is determined more rigorously than $V_p$ as mentioned above, this common behavior of $V_s$ must reflect an intrinsic property of the FIG. From the linear fit of the $V_s$ vs. $H$ plot shown in Figs. 4(g) – 4(i), we deduce g values as follows: 1.81 ± 0.43, 2.14 ± 0.22, and 1.96 ± 0.25 for the [001], [100], and [110] directions, respectively. Thus, our g-factor is isotropic within error bars and in good agreement with the g-value of 1.92 determined from the field-induced splitting of the neutron spin-resonance peak [72]. From the analysis of the temperature dependence of $H_{c2}$, Won et al. [73] reported an anisotropic g-factor: 1.5 for [001] and 0.62 for [100]. In a Pauli-limited superconductor, the critical field [74] is given by $H_P = \sqrt{2}\Delta(0)/g\mu_B$, where $\Delta(0)$ is the SC gap at H = 0. Using our $\Delta$ and g values, we estimate $H_P$ = 8.9 – 11.7 T, 6.2 T, and 6.7 T for [001], [100], and [110] directions, respectively. Note that the in-plane $H_P$ values are much smaller than the measured upper critical field, $H_{c2}$ = 11.8 T [31], warranting a revisit to the widely accepted Pauli-limited nature of the pairing in CeCoIn$_5$.

## V. DIRECT INVOLVEMENT OF LOCAL MOMENTS IN THE PAIRING PROCESS

The FIG in CeCoIn$_5$ is robust and reproducibly observed in multiple single crystals from different sources (Figs. S16, S17 in [39]) along all three crystallographic directions at $T < T_c$ and $T_c < T < T_p$, suggesting a common physical origin. Our conductance data on a (001) junction taken at two temperatures above $T_p$, namely, at $T$ = 10 K and 15 K, are shown in Fig. 5. Here, with increasing field, the broad ZBCP is only suppressed gradually without showing a clear signature for the FIG up to 14 T, eventually merging into the background. This distinct behavior above $T_p$ points to a concomitance of the pairing gap and the FIG, suggesting that the FIG is closely tied to the pairing mechanism. In addition, the FIG doesn't show any dependence on the field direction relative to the junction plane in all junctions (Fig. S18 in [39]). This is in line with the neutron spin resonance in CeCoIn$_5$ occurring at scattering wave vectors in three spatial dimensions [16].

As mentioned earlier, the compelling experimental fingerprint for spin fluctuation mediated pairing in the cuprate and Fe-based HTS is the spin resonance peak at $\omega = \Omega_{res}$ detected by inelastic neutron scattering, which also shows up in tunneling conductance as an additional dip-hump structure at $eV = \Delta + \Omega_{res}$ outside the coherence peaks [7-11] (Sect. 11 in [39]). The origin of the neutron resonance peak at $\Omega_{res} = 0.6$ meV in CeCoIn$_5$, despite the original interpretation as such a fingerprint [16], remains controversial [17,18], and the dip-hump structure is not observed in tunneling, neither in our PTS nor in the previous STS measurements [19-23]. Recently, van Dyke et al. [75] reproduced the neutron spin resonance peak by solving the SC gap equations and claimed the spin fluctuation mechanism in CeCoIn$_5$, but without accounting for the missing feature in tunneling conductance. It is clear that the pairing mechanism in CeCoIn$_5$ is yet to be determined. Below, we show that the signatures observed in our tunneling spectra are closely related to the pairing mechanism.

Coleman and coworkers [25,26] proposed a novel pairing mechanism based on the two-channel Kondo effect [27-29]. According to this theory, localized moments due to $4f^1$ electrons in CeCoIn$_5$ can be screened by conduction electrons via two channels. If this two-channel screening occurs cooperatively, the conduction electrons are effectively paired via the Kondo effect, leading to a composite pair. Here, the two-fold degeneracy of the crystal-field-split ground state Kramers doublet [59] is crucial, as it is for the single-channel Kondo effect. It is expected that, with the application of a magnetic field, the degeneracy is gradually lifted due to the Zeeman splitting, ultimately suppressing the composite pair formation. While a smoking gun evidence remains to be found, this exotic pairing has been invoked to explain the anomalous evolution of the gap structure observed in London penetration depth measurements on Ce$_{1-x}$Yb$_x$CoIn$_5$ [76,77].

As discussed earlier, our conductance spectra for non-nodal and nodal junctions exhibit distinct field evolutions. At low fields, the non-nodal junctions exhibit the pairing gap, whereas it is not seen in the nodal junction. This may be accounted for as being due to the sign-change of the $d_{x2-y2}$-wave order parameter, causing pairs to be broken on the nodal surface, as is well known for the high-$T_c$ cuprates [42-45]. However, stronger evidence for ABS in CeCoIn$_5$ is yet to be found since its characteristic signatures seem to appear on top of much stronger background, namely, a Kondo resonance over an energy scale comparable to $\Delta$, as mentioned earlier. This suggests that CeCoIn$_5$ may possess a more complex SC order parameter than a simple $d$-wave form. If the ZBCP arises from Kondo resonant tunneling, it would split under magnetic fields due to the Zeeman splitting of the Ce $4f^1$ moment. This is supported by the fact that the ZBCP begins to split at low fields in the absence of an apparent pairing gap. For non-nodal junctions, the FIG appears to be masked by the pairing gap until the field becomes strong enough to break a large portion of the pairs. These unpaired electrons can then participate in resonant and inelastic tunneling involving the localized moment [64,65], which is consistent with the $V_p$ undergoing a crossover at $H_{cr}$. Toward a more microscopic understanding of the crossover, two characteristic energy scales instead of nominal bias voltages should be compared, namely, $\Delta$ and $E_Z$. By solving the equation, $\Delta(H) = \Delta(0)\sqrt{1-(H/H_{c2})^2} = 1/2g\mu_B H$, and using our extracted values for $\Delta(0)$ and g along with the known $H_{c2}$, we obtain the crossing field, $H_c \sim 4.6$ T and 7.0 T for the (001) and (100) junctions, respectively, as marked by the gray line in Figs. 4(g) and 4(h). Notably, $H_c$ is anisotropic, as is $H_{c2}$ but unlike the isotropic $H_{cr}$, further supporting that the pairing mechanism and the FIG are intimately tied (as $H_{c2}$ depends on the depairing mechanism, orbital or Pauli-limited). While the pairing gap signature in the (001) junction is missing for $H > H_{cr}$ due to the closeness of $H_c$ to $H_{c2}$ in this direction, it is seen to coexist with the FIG in the (100) junction in some field range above $H_c$ since $H_c$ is much smaller than $H_{c2}$ in this direction (see Fig. S14 in [39]). At high temperature ($T > T_p \approx 5$ K), the pair formation might be suppressed presumably because increased thermal fluctuations weaken the cooperative effect between the two Kondo screening channels. Thus, the concomitance of the FIG with the pairing gap below $T_p$ might be due to the Kondo resonance itself playing a key role in the pair formation.

In the high-field limit, the FIG is of qualitatively similar V-shape among the non-nodal junctions, whereas it exhibits quite a different structure in the nodal junction, as shown in Fig. 6(a) and 6(b), respectively. These curves are compared with computed ones based on the AA theory [62,64,65] shown in Fig. 6(c). As mentioned earlier, the tunneling conductance involving Kondo impurities has been qualitatively accounted for by this theory [61-63]. Here, the conductance frequently exhibits U-shape in the high-field limit. This is because both the spin-flip inelastic tunneling and the Kondo resonant tunneling, whose conductance contribution is denoted as $G_2$ and $G_3$, respectively, in the literature, give rises to a step-like abrupt increase at bias voltages corresponding to $\pm E_Z$. Apparently, this is not the case for our data since the computed curves don't resemble them at all. On a close look, there exist two linear regions within the FIG for the (110) junction and the boundary appears to be close to $E_Z$. We associate the discrepancy in the FIG observed in between our tunneling data and the computed curves with the non-trivial nature of the Kondo resonance, which, in turn, is tied to the nature of pairing. The cooperative two-channel Kondo effect, proposed to give rise to the pairing in $CeCoIn_5$, may lead to such unusual Kondo resonance. Also, we note that theoretically the same effect could explain the non-Fermi liquid behavior [27] clearly observed in this compound below ~20 K [24,49,78], coincident with the onset temperature for the ZBC's upturn, as seen in Fig. 2(c). For a full account of the FIG, it is desirable to formulate a more microscopic model that explains, both qualitatively and quantitatively, how the magnetic field suppresses such pairing that directly involves the localized $f$-moments. Such model would also take into account the exact ground state for the Ce $4f^1$ electron in CeCoIn$_5$ as it has been recently identified to be a Kramer's doublet, that is, $\Gamma_{7-} = \alpha|\pm 5/2\rangle + \beta|\mp 3/2\rangle$, arising from the crystal-electric field effect [59].

## VI. CONCLUSION

Our PTS data on CeCoIn$_5$ and analyses reveal the existence of preformed pairs below $T_p \sim 5$ K, well above $T_c = 2.3$ K, consistent with the previously reported STS and several bulk measurements. Upon lowering the temperature below $T_p$, both the density of pairs and their lifetime increase due to the reduction in thermal fluctuations, allowing them to condense into a phase coherent state at $T_c$. The pairing symmetry inferred from the directional dependence is $d_{x2-y2}$, in overall agreement with the literature, although its detailed nature is yet to be unraveled. With the application of a magnetic field, the pairing gap gradually turns into the FIG. And the FIG appears only at temperatures up to where the pairing gap persists. This concomitance of the pairing gap and the FIG provides a clue for the microscopic pairing mechanism in CeCoIn$_5$. The FIG exhibits linear field dependence and non-trivial structure, suggesting that the pairing in CeCoIn$_5$ may directly involve localized moments, e.g., via the cooperative two-channel Kondo effect that has been proposed theoretically.

**ACKNOWLEDGEMENTS**

We acknowledge fruitful discussions with A. V. Balatsky, A. V. Chubukov, P. Coleman, P. Ghaemi, J. D. Thompson, and K. Yang. The work at FSU (KS, SZ, LHG, and WKP) was supported by the US National Science Foundation (NSF), Division of Materials Research (DMR), under Award No. NSF/DMR-1704712. YL and RB acknowledge support from Department of Energy through the Center for Actinide Science (an EFRC funded under Award DE-SC0016568). A portion of this work was performed at the National High Magnetic Field Laboratory, which is supported by the NSF Cooperative Agreement No. NSF/DMR-1644779 and the State of Florida. The work at UCSD was supported by the US DOE-BES, DMSE, under Grant No. DEFG02-04-ER46105 (single crystal growth) and US NSF/DMR-1810310 (physical properties measurements).

**FIGURE CAPTIONS**

**Fig. 1. Comparison of tunneling conductance along three major crystallographic directions of CeCoIn$_5$: (a) (001), (b) (100), and (c) (110).** The temperature is 20 mK and the applied magnetic field is 0.2 T (Pb driven normal). Main panels show the normalized conductance, $G_b(V)$, obtained by dividing out the raw data with the (~ parabolic) background, as shown in the right insets. The lines in (a) & (b) are best fits to the *d*-wave BTK model, with fit parameters ($\Delta$, $\Gamma$, Z) = (0.66 meV, 0.042 meV, 2.28) and (0.535 meV, 0.198 meV, 1.21), respectively, where $\Gamma$ is the quasiparticle lifetime broadening parameter and Z represents the dimensionless barrier strength. The red solid line in (c) is a fit of the wider peak of Lorentzian shape to the Frota function depicting a Kondo resonance (Ref. 45): $G_{\text{fit}}(V) = 0.985 + 0.18 \times Re\sqrt{\left(\frac{0.75 \times 10^{-3} i}{V + 0.75 \times 10^{-3} i}\right)}$. The left inset is the conductance at low bias further normalized by the Frota fit background shown in the main panel.

**Fig. 2. Temperature evolution of the background-normalized tunneling conductance in CeCoIn$_5$.** The applied magnetic field is kept at 0.2 T. (a) – (c), waterfall plots of the conductance at varying temperature for (001), (100), and (110) junctions, respectively. Curves are shifted vertically in (a) and (b) for clarity. (d) – (f), corresponding color contour plots of the conductance with the y-axis (temperature) in logarithmic scale. The right insets show temperature dependence of the zero-bias conductance. In the left inset of (a), conductance curves around $T_p$ ($\approx$ 5 K) are plotted to show more clearly the evolution from a broad ZBCP to gaplike split peaks. The white horizontal dashed lines are to mark the bulk $T_c$ (2.3 K).

**Fig. 3. Opening of the pairing gap well above $T_c$ in CeCoIn$_5$.** (a) Temperature-dependent $G_n(V)$ for a (001) junction in which the pairing gap emerges out of a ZBCD below $T_p$ instead of a ZBCP. The coherence peaks become sharp at low temperature. The curves overlap well at high bias. Inset: Magnified view of the gap edge. (b) $G_b(V)$ curves (black symbols) and their best fits (solid orange lines) to the d-wave BTK model. $G_b(V)$ is obtained by dividing out each $G_n(V)$ with $G_n(V)$ at 5 K. (c) Best-fit values for $\Delta$ and $\Gamma$. Z is kept to be a constant, 5. At 0.4 K, ($\Delta$, $\Gamma$, Z) = (0.855 meV, 0.179 meV, 5.0). $\Delta$ remains finite above $T_c$ and extrapolates to zero at $T_p \sim 5$ K, as indicated by the dashed line.

**Fig. 4. Magnetic field evolution of the background-normalized conductance in CeCoIn$_5$.** (a) – (c), waterfall plots of the conductance for varying magnetic field applied along the junction normal at temperatures well below $T_c$ for (001), (100), and (110) junctions, respectively. (d) – (f), the same at temperatures well above $T_c$. The (100) and (110) junctions are the same ones as shown in Fig. 1. Curves are shifted vertically in (a), (b), (d), and (e) for clarity. (g) - (i) Peak position, $V_p$ ((filled green circles), and steepest slope point, $V_s$ (filled black squares), of the FIG at low temperature (Ref. 33, Sect. 9). In (g) & (h),

Δ is also plotted for $H = 0.2$ T with $(\Delta, \Gamma, Z) = (0.69$ meV, $0.405$ meV, $1.81)$ for (g) and $(0.535$ meV, $0.198$ meV, $1.21)$ for (h). The crossing field ($H_c$) is indicated by gray lines. In (h), slope-changing points due to the pairing gap, $V_{ps}$, above $H_c$ are also shown by filled triangles. Dash-dotted and dashed lines are linear fits to $V_p$ and $V_s$, respectively. The blue dotted lines crossing the point $\Delta(0.2$ T$)$ in (g) and (h) show field dependence of the pairing gap according to the GL theory (see the text).

**Fig. 5. Absence of the FIG above $T_p$.** Magnetic field dependence of the high-bias normalized conductance for a (001) junction on CeCoIn$_5$ at (a) $T = 10$ K and (b) $T = 15$ K. The broad peak at zero bias is suppressed gradually with increasing magnetic field, merging into the background without the FIG feature. Insets: $G_b$ curves showing a very small change with the field. The faint gaplike feature appearing at high field (zero-bias conductance depth smaller than 0.5% for 14 T) is unlikely to be intrinsic as it depends on the background normalization, e.g., the bias range taken for the quasi-linear background conductance.

**Fig. 6. Comparison of the FIG in CeCoIn$_5$ with calculation based on the Anderson-Appelbaum (AA) theory.** (a) & (b), Experimental $G_b$ curves for the (100) and (110) junctions, respectively, taken at 20 mK and two fields in the high field limit where the FIG is most pronounced. (c) $G_n$ curves calculated based on the AA theory for the same fields and temperature as in (a) & (b). The expressions for $G_2$ (spin flip inelastic tunneling) and $G_3$ (Kondo resonant tunneling) terms are adopted from Ref. 61 with the weight factor of 0.5 per each. The g-factor is 2 and the spin is 1/2. The vertical gray lines indicate the bias voltages corresponding to $\pm E_Z$ at $H = 18$ T.

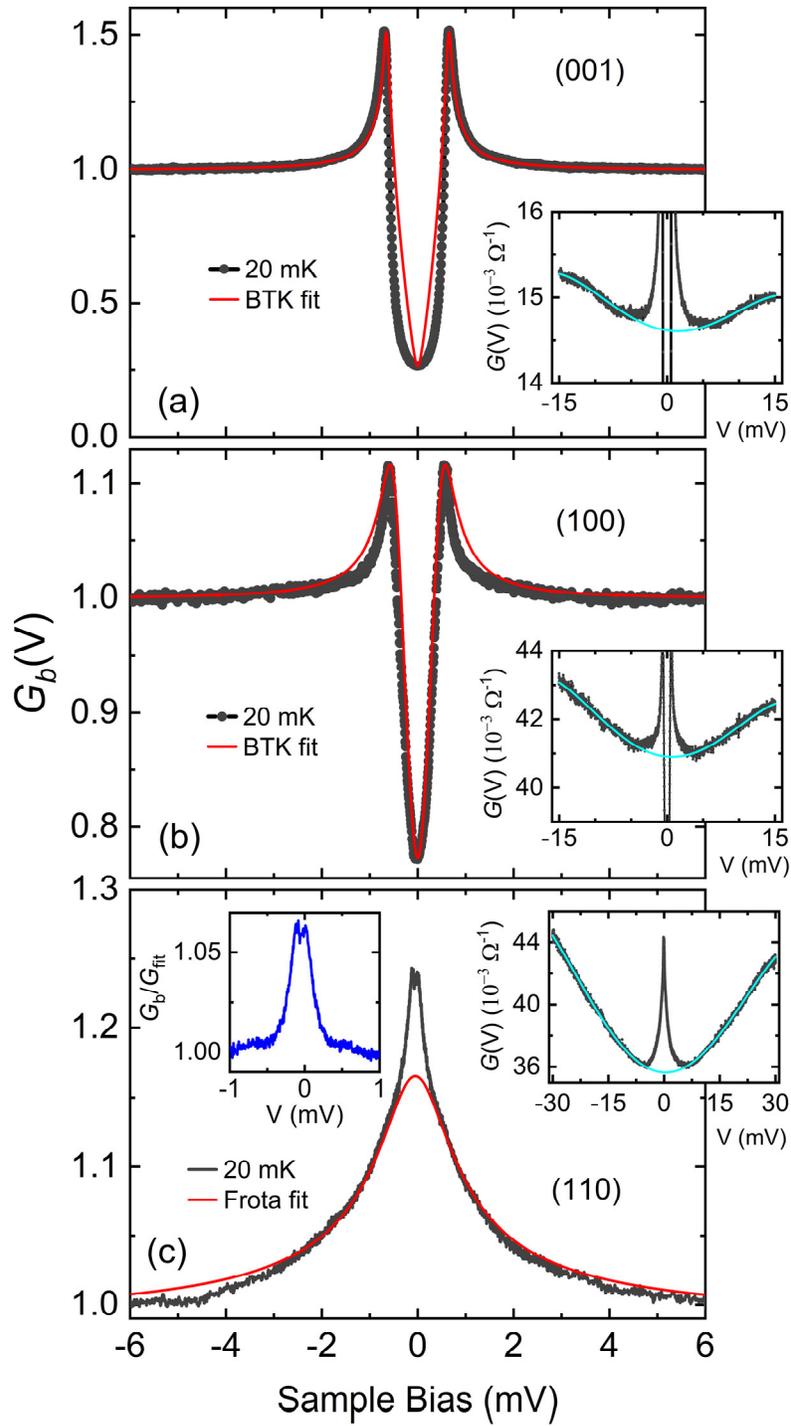

Figure 1, K. Shrestha et al.

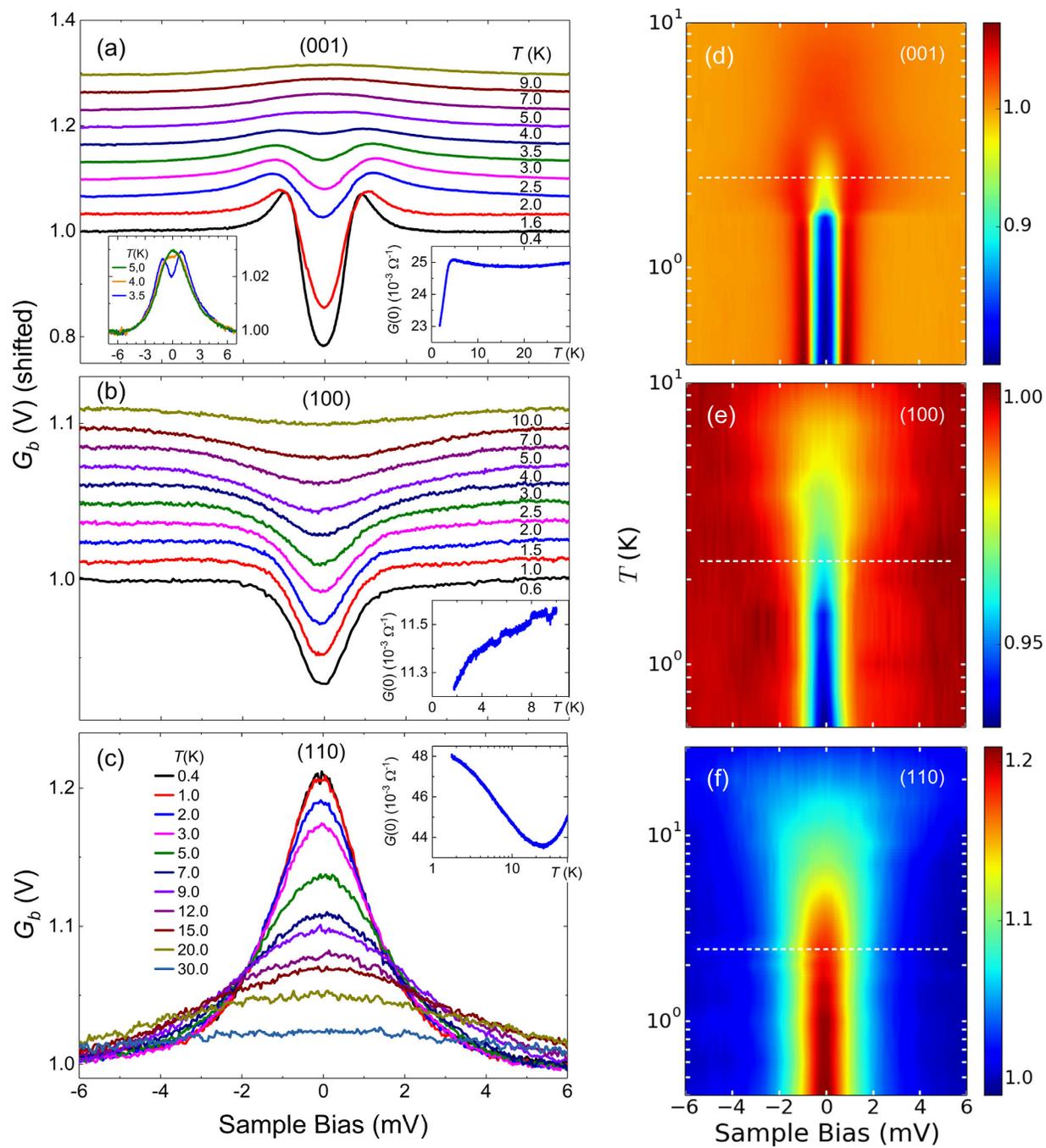

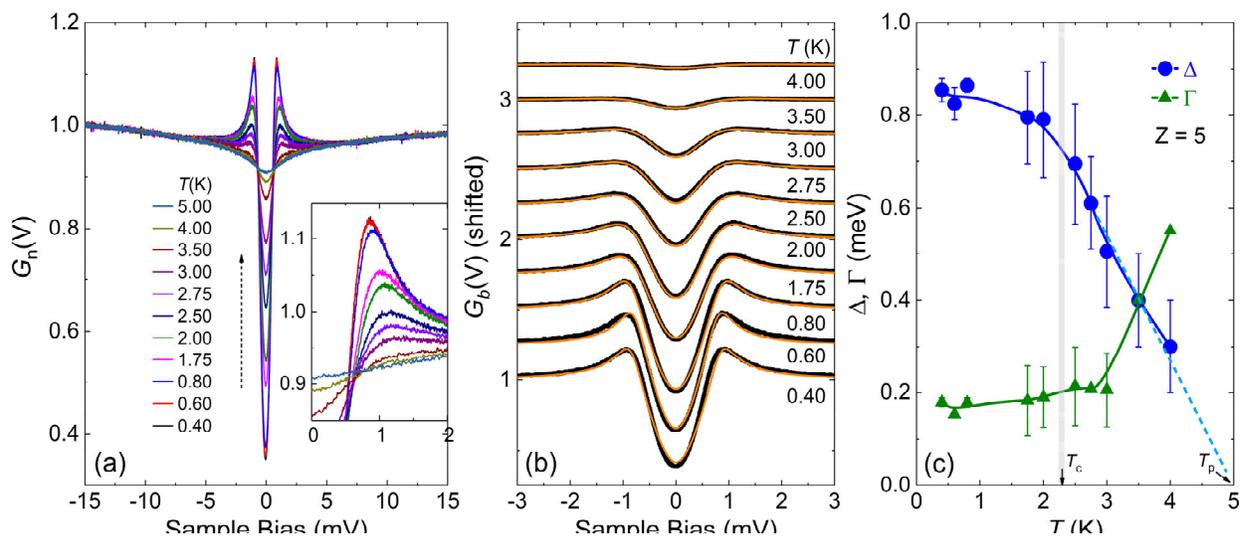

Figure 3, K. Shrestha et al.

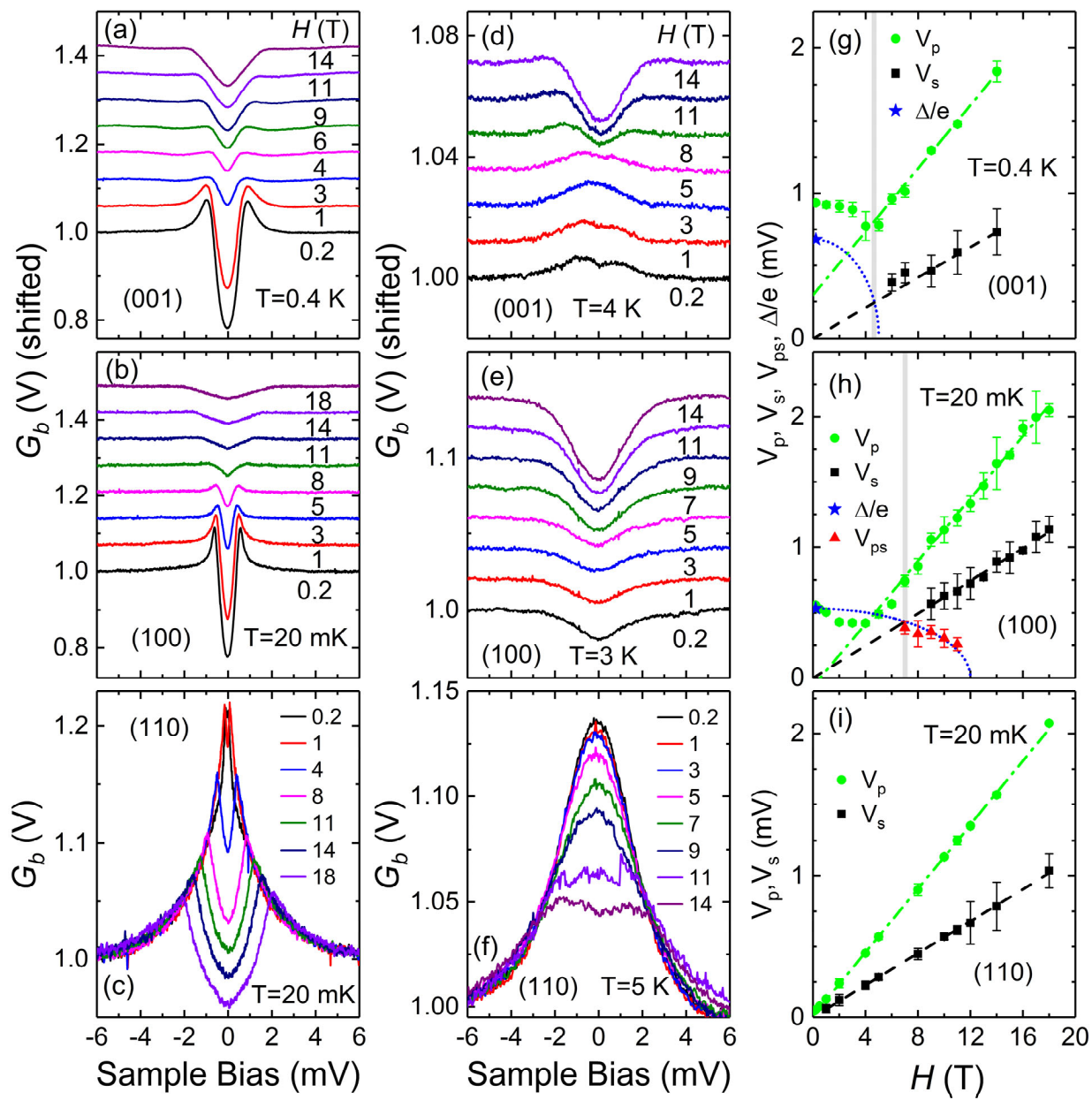

Figure 4, K. Shrestha et al.

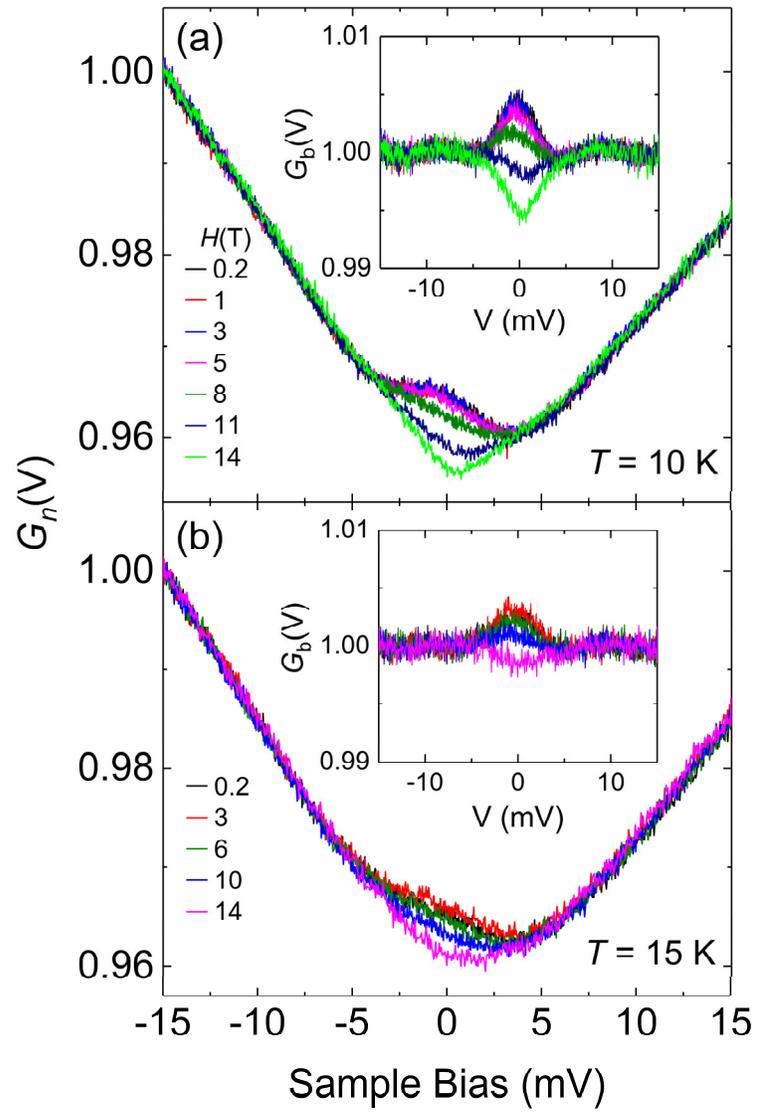

Figure 5, K. Shrestha et al.

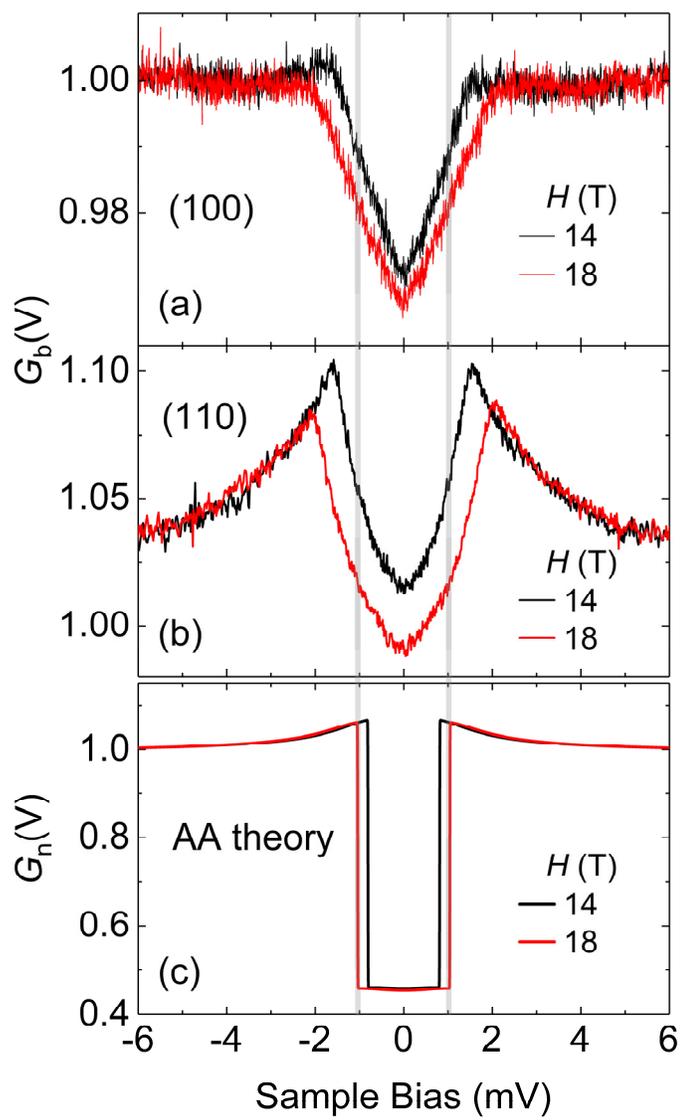

Figure 6, K. Shrestha et al.

**Supplemental Material for**

**Spectroscopic Evidence for the Direct Involvement of Local Moments in the Pairing Process of the Heavy fermion Superconductor CeCoIn$_5$**

K. Shrestha[1,†], S. Zhang[1,2], L. H. Greene[1,2], Y. Lai[1,2], R. E. Baumbach[1,2], K. Sasmal[3], M. B. Maple[3], and W. K. Park[1,*]

[1]*National High Magnetic Field Laboratory, Florida State University, Florida, 32310, USA*
[2]*Department of Physics, Florida State University, Florida, 32306, USA*
[3]*Department of Physics, University of California, San Diego, California, 92093, USA*

**Table of Contents**

[†]Present address: Department of Chemistry and Physics, West Texas A&M University, 2501 4th Ave, Canyon, Texas 79016, USA.

[*]To whom correspondence should be addressed. Email: wkpark@magnet.fsu.edu.

# 1. Materials and Methods

## 1.1. Growth of CeCoIn$_5$ single crystals

High-quality CeCoIn$_5$ single crystals were grown using the molten metal-flux growth technique with an excess of indium as flux [1]. High purity elements of *Ce* (Alfa Assar, >99.9%), *Co* (Alfa Assar, 99.998%) and *In* (Alfa Assar, 99.9999%) were loaded inside cleaned and dried alumina crucible with the mixture of constituents sealed in evacuated quartz tubes. The quartz ampules were heated to 1100 °C and then slowly cooled to 600 °C, at which point the molten indium flux was finally decanted with a centrifuge. Optimal CeCoIn$_5$ single crystals were obtained under these conditions. Grown crystals were plate-like rectangular shaped, clean, and shiny with typical dimensions up to 3.5×3×1.0 mm$^3$.

## 1.2. Polishing the crystal surface

Shiny pieces of CeConIn$_5$ crystals (size ≈ 2 × 1 mm$^2$) were selected and fixed on disc-shaped Stycast® epoxy (2850-FT) molds. For the nodal surface, the crystal was cut along [110] direction, and its orientation was confirmed with a single crystal x-ray diffractometer before fixing it to the epoxy disc. Since the surface roughness is one of the important factors for high-quality tunnel junction, we polished the crystal surface to mirrorlike shininess first using a diamond lapping film (particle size = 1 µm) and then non-sticky silica colloid (particle size = 0.04 µm). The sample was pressed manually against the lapping film and isopropyl alcohol was used as a lubricant. The polished surface was inspected under an optical microscope from time to time to check the status. Once uniform roughness is achieved, we switched to the next step, in which the sample was rubbed against a polishing cloth using colloid as a lubricant. Residual colloid particles on the surface were cleaned by ultrasonication in acetone, isopropanol, and then methanol for 1 minute in each solution. Atomic force microscope (AFM) was used to make sure the polished surface is free of colloidal particles and also to measure the smoothness. Figure S1 (a) shows an AFM image of the (001) surface. The surface looks uniform and smooth with an overall peak-to-dip distance of ~1 nm, as shown in Fig. S1 (b). Similar smoothness was also observed in other crystallographic surfaces. This roughness is good enough to proceed for the next step of junction fabrication process.

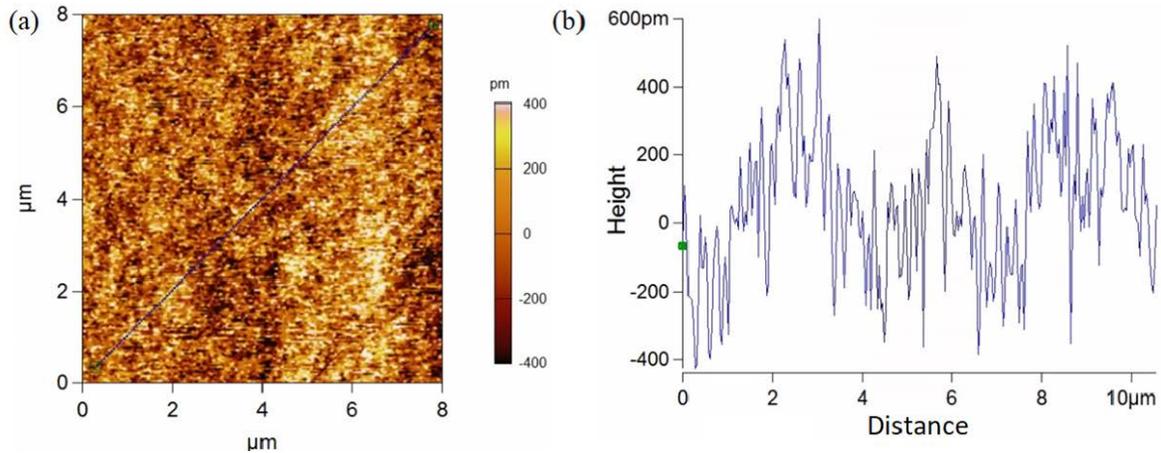

**Figure S1. Surface analysis of the polished crystal.** (a) AFM image of the polished (001) surface of a CeCoIn$_5$ crystal. The color contrast reflects peak and dip areas on the surface. (b) The surface roughness along the solid black line shown in (a). The average peak-to-dip distance over a 12 µm line scan is less than 1 nm.

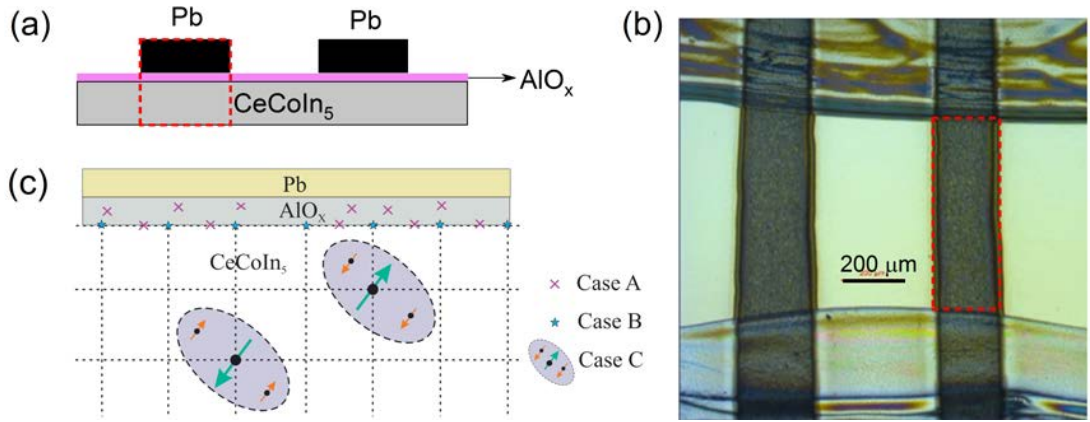

**Figure S2. Structure of a CeCoIn$_5$/AlO$_x$/Pb tunnel junction**: (a) Schematic cross-sectional view of a CeCoIn$_5$/AlO$_x$/Pb tunnel junction. A thin layer of Al (2 – 2.5 nm) is deposited onto the polished surface and subsequently oxidized to form insulating AlO$_x$. Then, Pb thin film (~ 250 nm) is deposited on top of the AlO$_x$ layer as a counter-electrode. The dashed rectangle indicates the region where the junction is formed. (b) An optical microscope image of the junction. Duco cement was painted along the crystal edges (top & bottom) and the Pb thin film (dark rectangular strips) was deposited through a shadow mask. The dashed rectangle represents the junction area and the bright area is where the crystal is coated only with AlO$_x$. (c) Schematic drawing of the junction structure to illustrate three possible origins for the FIG. Case A: extrinsic magnetic moments induced in the AlO$_x$ barrier or at the interface. Case B: magnetic moments due to Ce$^{3+}$ ions at the *surface* of the CeCoIn$_5$ crystal. Case C: *bulk* effect related to pairing involving localized Ce $4f^1$ moments.

**1.3. Preparation of CeCoIn$_5$/AlO$_x$/Pb tunnel junctions**

A polished CeCoIn$_5$ crystal was loaded into a high vacuum chamber, which was pumped down to a base pressure of ~4 × 10$^{-7}$ Torr. First, the sample surface was cleaned using a low-energy (100 eV) Ar ion beam. This process removes residual contaminants on the surface generated from the polishing process. To avoid the formation of micro-shorts between the top and bottom electrodes in a junction, the thickness of the tunnel barrier should be larger than the roughness of the sample. Thus, an Al layer of thickness 2 – 2.5 nm was deposited onto the sample surface using DC magnetron sputtering, which was subsequently oxidized to form AlO$_x$. The oxidation was done with an oxygen plasma generated by DC glow discharge. The typical time interval between the deposition and oxidation of Al is less than 3 minutes. The sample was then removed from the chamber for depositing Pb as a counter-electrode. Before this, the crystal edges were painted with a diluted Duco Cement to define the junction area and to prevent the junction from being electrically shorted. The Pb deposition process was carried out in a thermal evaporator in which thermally evaporated Pb vapors pass through a shadow mask to form narrow strips on the sample surface. Figure S2 shows (a) a schematic cross-sectional view of the junction and (b) an optical image of the junction in top. The dashed rectangle in both plots denotes the region where the junction is formed.

**1.4. Junction characterization and junction quality**

Differential conductance of the tunnel junction was measured using a standard four-probe lock-in technique in a Quantum Design 14 T PPMS (physical property measurement system) equipped with He-3 option. Measurements at lower temperatures down to $T = 20$ mK and under higher fields up to 18 T were carried out in the Millikelvin Facility at the National High Magnetic Field Laboratory, Tallahassee, Florida.

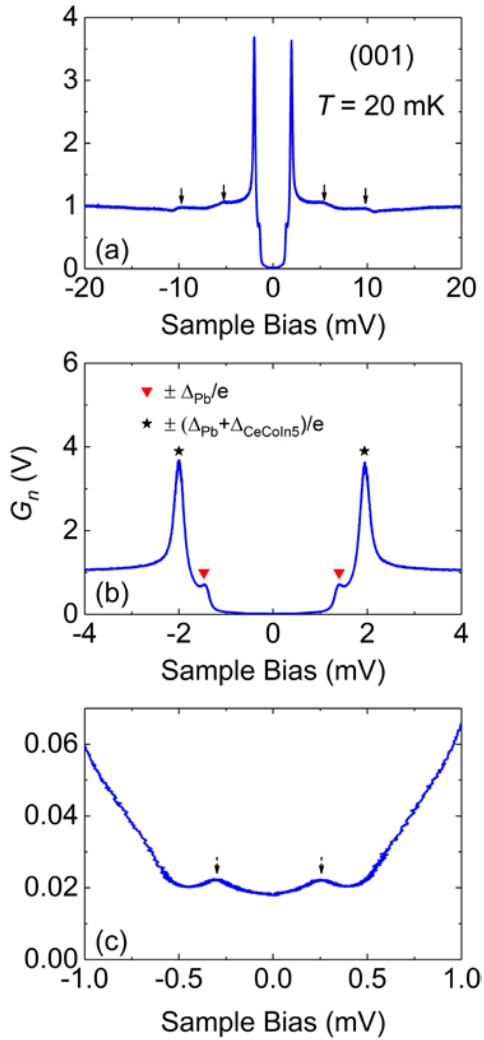

**Figure S3. Normalized tunneling conductance across a (001) CeCoIn$_5$/AlO$_x$/Pb junction.** The measurement temperature is 20 mK and no magnetic field is applied. (a) In addition to sharp coherence peaks, the conductance spectrum shows well-defined phonon features (around ±5 mV and ±10 mV, see the arrows) and the zero bias conductance is nearly zero. In the low bias region, there are multiple peaks corresponding to $\pm(\Delta_{Pb} + \Delta_{CeCoIn5})$ at an averaged bias voltage of 1.98 mV, $\pm\Delta_{Pb}$ at 1.44 mV, and at about 0.25 mV as shown by the arrows in (b) and (c).

Normalized conductance of the junctions on (001), (100), and (110) surfaces are shown in Figs. S3, S4, and S5, respectively. Since the measurements were made at zero field, the conductance is expected to show structures for an S-I-S' type tunnel junction. For the [001] junction, there exist multiple peaks corresponding to $\Delta_{Pb}$ (1.38 meV) + $\Delta_{CeCoIn5}$ (0.65 meV), $\Delta_{Pb}$ (1.38 meV) and at about 0.25 meV. The [100] junction also exhibits similar features as in the (001) junction; however, the peak corresponding to $\Delta_{Pb}$ is sharper than the one at $\Delta_{Pb} + \Delta_{CeCoIn5}$, opposite to the (001) junction, and the peak at 0.25 meV is absent. The nodal junction shows the Pb SC gap features but no signatures for $\Delta_{CeCoIn5}$. This is presumably because Cooper pairs broken on the nodal surface of the CeCoIn$_5$ crystal. The nearly-zero conductance at zero bias that is seen in all three junctions indicates that they fall to the tunneling limit. Thus, all the structures described above must be due to quasiparticle tunneling between two superconductors (S & S'). However, unlike in a tunnel junction between two conventional superconductors, some structures can't be explained by a tunneling model for a simple S-I-S' junction. Their exact origins remains to be investigated in the future. At any rate, the sharp conductance features observed in all junctions on three crystallographic surfaces attest the junctions' high quality.

The conductance data that reflect only the superconductivity in CeCoIn$_5$, which are presented in the main text, were obtained by driving the Pb normal with a small magnetic field (H = 0.2 T). Unless otherwise specified, througout the entire paper, the magnetic field was applied parallel to the identified crystallographic orientation of the crystal surface, that is, perpendicular to the junction plane.

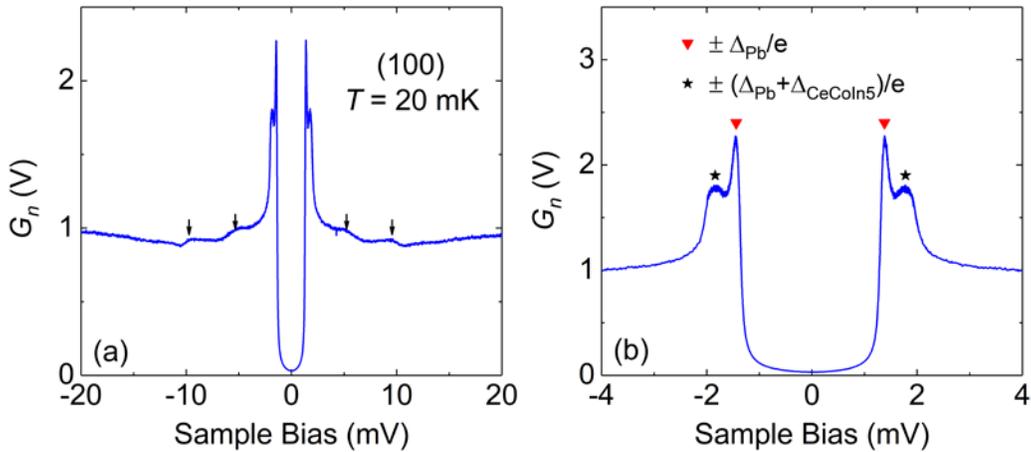

**Figure S4. Normalized tunneling conductance from a (100) CeCoIn$_5$/AlO$_x$/Pb junction.** The measurement temperature is 20 mK and no magnetic field is applied. (a) The Pb phonon features are clearly observed (around ±5 mV and ±10 mV, see the arrows) and the zero-bias conductance is close to zero. There are two gap features at an averaged bias voltage of 1.81 mV corresponding to $\pm(\Delta_{Pb} + \Delta_{CeCoIn5})$ and at 1.42 mV for $\pm\Delta_{Pb}$, as indicated by the arrows in (b). Unlike the (001) junction, the $\Delta_{Pb}$ peak is more pronounced than the peak at $\Delta_{Pb} + \Delta_{CeCoIn5}$.

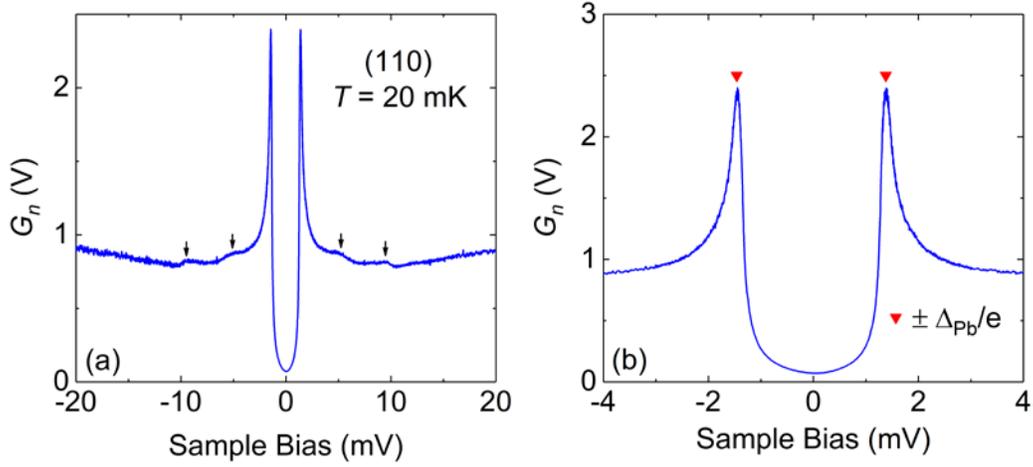

**Figure S5. Normalized tunneling conductance across a (110) CeCoIn$_5$/AlO$_x$/Pb junction.** The measurement temperature is 20 mK and no magnetic field is applied. (a) The conductance spectrum shows well-defined phonon features (around ±5 mV and ±10 mV, see the arrows), however, the zero-bias conductance is not as low as in (001) and (100) junctions. There exist peaks corresponding to ±Δ$_{Pb}$ at an averaged bias voltage of 1.42 mV but no peaks corresponding to ±(Δ$_{Pb}$ + Δ$_{CeCoIn5}$) are seen, unlike the (001) and (100) junctions. This could be because the pairs are broken on the nodal surface of CeCoIn$_5$.

## 2. BTK analysis and tunneling cone effect

To extract the SC gap, we analyzed conductance spectra using the *d*-wave Blonder-Tinkham-Klapwijk (BTK) theory [2]. This was done using a MATLAB code that automatically finds optimal values for three fitting parameters Δ, Γ, and Z by adjusting them such that the computed curve matches with the data in the location of the coherence peak (highest point), in its height, and in the depth of the zero-bias conductance dip (lowest point), respectively. After the best-fit curve is found, variances around both zero-bias and the coherence peak are calculated to be used as references in determining the error bars. For Δ, Δ is slowly increased by increment of 0.005 meV starting from the best-fit value. Within each loop, Γ is tuned in a similar way as described above to find a matching curve. The loop stops when the above-mentioned variances computed for both regions become twice the variances in the best-fit case, or once the loop has repeated a thousand times. The error bar for Γ was determined in a similar manner by tuning Δ in each loop.

If the momentum of tunneling electrons spans over the entire hemisphere of the Fermi surface, the computed tunneling conductance exhibits V-shape, characteristic of the single particle density of states of a d-wave superconductor, as shown in Fig. S6 for the range of angular integration θ = π/2. This is the case for a (100) junction shown in Fig. 1(b). However, the conductance curve for a (001) junction shown in Fig. 1(a) is of U-shape. This could be due to the tunneling cone being narrow in this junction, as demonstrated in Fig. S6.

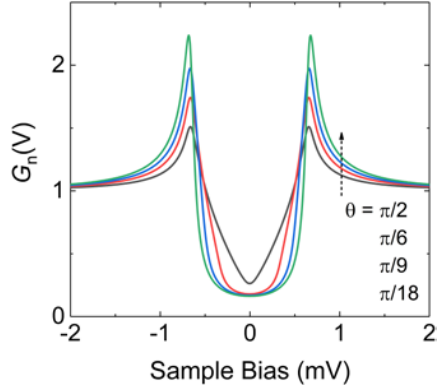

**Figure S6. Tunneling cone effect.** The tunneling conductance is calculated based on the $d$-wave BTK theory for a junction whose normal is along the lobe direction of the d-wave gap as a function of the range of angular integration, θ. The parameters from Fig. 1(a) are adopted: $T = 20$ mK, $\Delta = 0.66$ meV, $\Gamma = 0.042$ meV, and $Z = 2.3$.

### 3. Signature for pairing above $T_c$ in DC magnetic susceptibility

The temperature evolution of our planar tunneling data reveals an opening of the pairing gap at $T_p = 5$ K, well above $T_c$. Bulk properties such as thermal conductivity [3] and resistivity [4] also have shown signatures for a pseudogap in CeCoIn$_5$. Seeking further evidence on this, we have measured the temperature dependence of another bulk property, dc magnetic susceptibility ($\chi$). Figure S7 shows the $\chi$ vs. $1/T$ plot of CeCoIn$_5$ in the low-temperature region (below 15 K) with an applied field along the c-axis (a) and ab-plane (c). $\chi$ follows the Curie-Weiss $1/T$ behavior at high temperature, and then shows another $1/T$ dependence at lower temperatures. Both in-plane and out of plane $\chi(T)$ show a noticeable slope change near 5 K, as indicated by the dashed lines and also seen in the $d\chi/dT$ vs. T plot (Figs. S7b & S7d). This temperature is consistent with $T_p \sim 5$ K in our tunneling data. As the magnetic susceptibility in CeCoIn$_5$ originates from the $Ce^{3+}$ ions, the decrease in slope below $T_p$ can be interpreted as due to Ce $4f^1$ electrons being screened, possibly as they get involved in the pairing, as discussed in the main text.

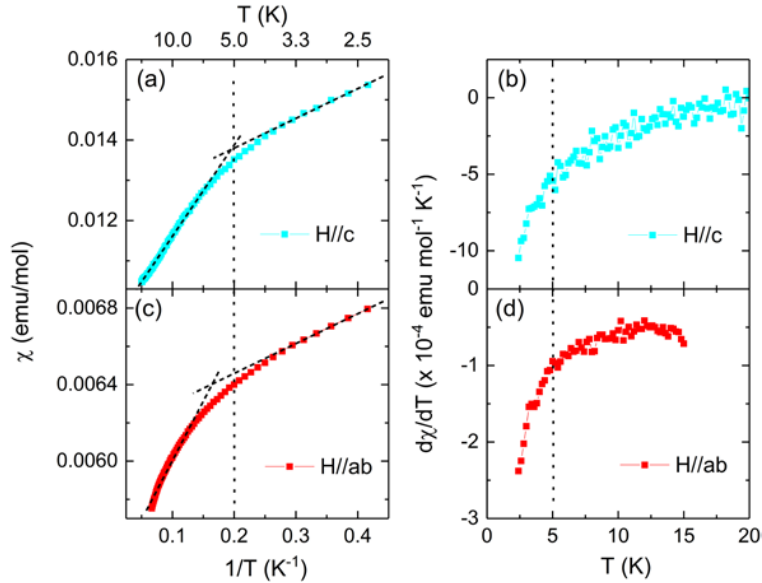

**Figure S7. Temperature evolution of the DC magnetic susceptibility ($\chi$).** The $\chi$ vs $1/T$ plot of CeCoIn$_5$ in the low-temperature region with the applied field ($H = 1000$ Oe) along the c-axis (a) and ab-plane (c). Top axes in both plots represent the temperature. Both the in-plane and out of plane $\chi(T)$ show a clear decrease in the slope below $T = 5$ K $\sim T_p$, as shown by the dashed lines and also in the $d\chi/dT$ vs. T plots (b), & (d).

## 4. Temperature and field dependent variation among junctions

We observed that the pairing gap emerges out of a broad peak in some junctions (Fig. 2 (a)) and a broad dip in other junctions (Fig. 2 (b)). Also, some junctions show a pronounced field-induced gaplike feature (FIG), whereas others do not (Fig. 2(b) and Fig. S9). A possible origin for this discrepancy in junction behavior may be the variation in Fano interference. Theoretically, there are two channels for electrons to tunnel into, the conduction band and the localized (or renormalized) $f$-band. These two channels interfere, known as Fano interference, with the Fano parameter ($q_F$) given as $q_F = (t_f \mathcal{V})/(t_c W)$, where $\mathcal{V}$ is the hybridization amplitude, W is the width of the Kondo resonance, $t_f$ and $t_c$ are matrix amplitudes for tunneling into the f-orbital and conduction band, respectively [5,6]. The conductance shape is determined by the $q_F$ value; A broad peak (or a dip) for a large (or small) $q_F$ value, as shown in Fig. S8. Empirically, junctions showing a pronounced FIG feature (Zeeman splitting of the 4*f*-level and consequent Anderson-Appelbaum(AA)-like conductance) have a broad peak at high temperature (Fig.2 (a) and Fig. S11). This means a large $q_F$ value, namely, predominant tunneling into the f-band and, thus, a more pronounced field effect. The junctions showing less pronounced field effect have a dip at high temperature (Fig. 2(b)), indicating small $q_F$, thus, predominant tunneling into the conduction band, which explains both the weaker FIG and pronounced SC gap features such as sharper coherence peaks and smaller zero-bias conductance (ZBC) (Fig. S9) since it is the conduction electrons that are paired after all.

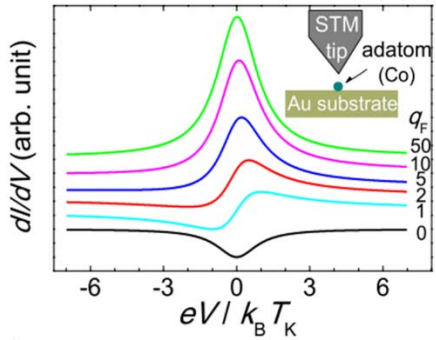

**Figure S8. Variation of the conductance shape depending on the Fano parameter ($q_F$).** Simulated conductance spectra for a Kondo impurity at different $q_F$ values. The conductance shows a zero-bias dip (anti-resonance) at $q_F = 0$, gradually turning into an asymmetric curve for intermediate $q_F$, and then into a zero-bias peak (resonance) at higher $q_F$ value due to Kondo resonant tunneling. Adapted from Ref. 6.

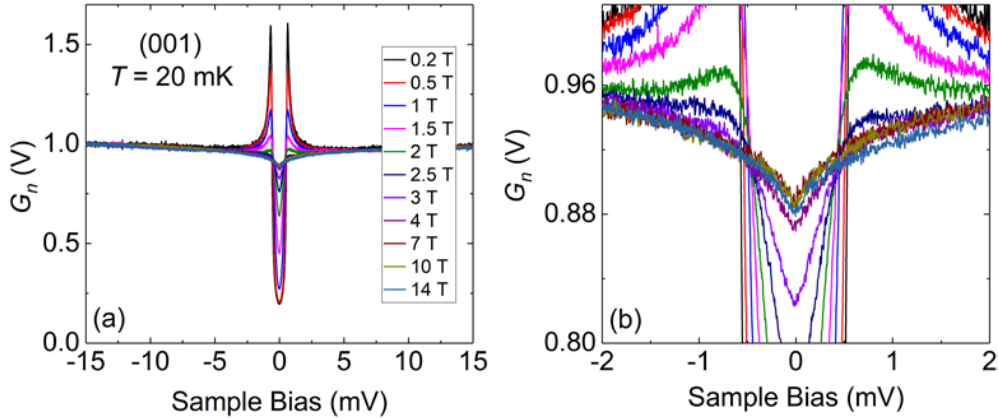

**Figure S9. Field evolution of a junction with small $q_F$ value.** (a) Normalized conductance for a (001) junction at 20 mK as a function of magnetic field. The applied field gradually suppresses the SC gap up to 4 T and then induces the FIG. The FIG does not increase with further increase of field, as shown in an enlarged view in (b). This weaker feature could be due to small $q_F$ for this junction so that tunneling into the conduction band is predominant than into the renormalized $f$-band.

## 5. Origin of the ZBCP – Kondo resonance

To understand the origin of the broad zero-bias conductance peak (ZBCP) that has been observed in all three directions. Since the ZBCP is most pronounced in (110) junctions, we traced its temperature evolution on a (110) junction from 1.75 K up to 50 K, as shown in Fig. S10(a). The conductance curve is parabolic at 50 K, as expected for a simple tunnel junction. From an analysis of this curve using a theoretical model

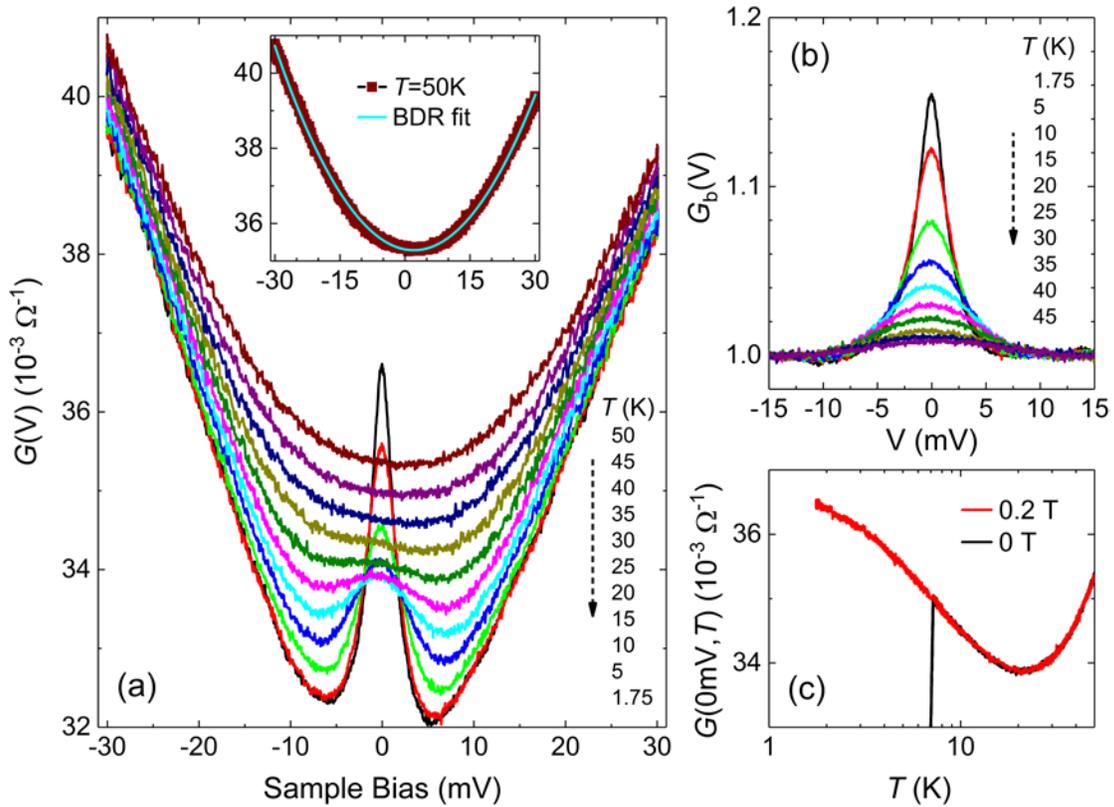

**Figure S10. Temperature evolution of the ZBCP.** (a) Temperature-dependent conductance spectra from a nodal junction. Inset: Barrier analysis of the 50 K data using the Brinkman-Dynes-Rowell (BDR) model. (b) The same conductance in (a) normalized by dividing out with the background at each temperature. A broad ZBCP develops beginning at T = 45 K and becomes sharper and narrower with decreasing temperature. (c) Zero-bias conductance as a function of temperature in log scale.

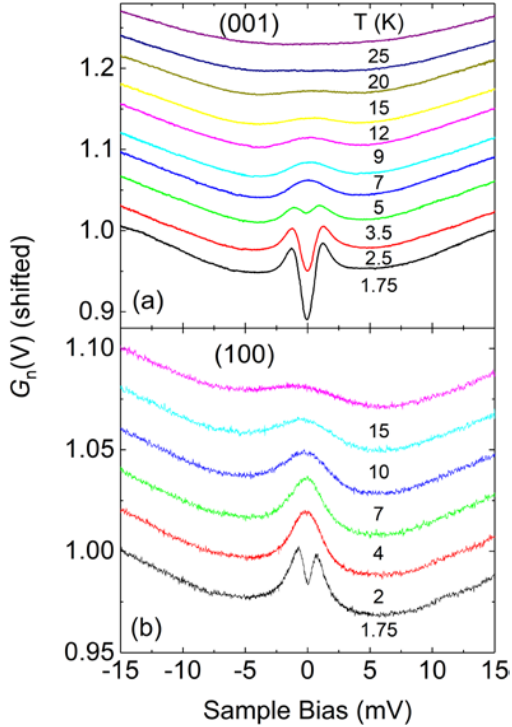

**Figure S11. Temperature dependence of the conductance along [001] and [100].** Raw conductance spectra for (a) (001) and (b) (100) junctions over a wide temperature range. The curves are shifted vertically for clarity. In both junctions, a broad peak at zero bias appears below T = 20 K and becomes sharper as the temperature is lowered. Compared to the (001) junction, the (100) junction does not show a clear pairing gap feature (double peak) at T = 2 K, namely, even below $T_c$, presumably due to the much larger smearing effect in this junction.

[7], as shown in the inset of Fig. S10(a), we obtain parameters characterizing the tunnel barrier: barrier height = 1.78 eV, barrier asymmetry = 0.38 eV, and barrier thickness = 22.57 Å, the latter being in good agreement with the intended thickness. With decreasing temperature, the conductance shape deviates from a parabola starting at T = 45 K, coincident with the coherence temperature ($T_{coh}$) at which heavy fermions start to form in CeCoIn$_5$ [8]. With lowering the temperature further, the ZBCP becomes narrower and sharper, more clearly observed in background-normalized curves shown in Fig. S10(b). The growth of ZBCP is also seen more quantitatively in the ZBC vs. temperature plot shown in Fig. S10(c). The logarithmic temperature dependence suggests that the ZBCP originates from a Kondo resonant tunneling. Similar broad peaks are also observed in other directions, [001] and [100] (Fig. S11). The discrepancy in the exact temperature below which the ZBCP develops (or even a dip rather than a peak) in junctions along different crystallographic directions could be either (i) intrinsic as it may reflect the anisotropy of hybridization process due to the anisotropic ground state orbital configuration [9] or (ii) rather junction specific (e.g., variation in the Fano parameter depending on the microstructural properties of the polished CeCoIn$_5$ surface). Further measurements and analysis are desired to nail down the exact origin.

## 6. Pairing gap to FIG crossover

As seen in the main text (Fig. 4), the tunneling conductance across non-nodal junctions shows a crossover behavior around 4 T from the pairing gap to the FIG. The measurement of the ZBC as a function of field reveals more details, as shown in Fig. S12 for a (001) junction. If the AA tunneling channel exists in addition to the superconductive tunneling (independently), the ZBC is expected to increase more slowly. A quantitative analysis of the ZBC data by considering different scenarios may bring further insights on the origin of the FIG.

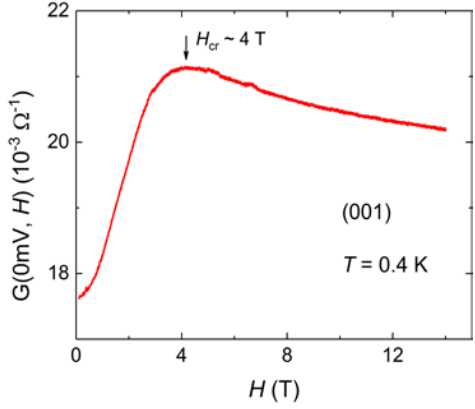

**Figure S12. ZBC vs H**. The zero-bias conductance is plotted as a function of magnetic field for a (001) junction. Similarly to the $G_b$ curves shown in Fig. 4, the ZBC undergoes a crossover at ~ 4 T.

## 7. Analysis of the ZBCP based on the single impurity Kondo resonance model

During the procedure for preparing a tunnel junction, extrinsic magnetic moments can be induced/introduced inside the tunnel barrier or at the interface. They may act like a single Kondo impurity, causing a ZBCP due to the Kondo resonant tunneling. A telltale signature of the ZBCP of such origin is its splitting under magnetic fields due to the Zeeman effect [10-12]. However, such single Kondo impurities, whether due to extrinsic moments or surface $Ce^{3+}$ ions, can be ruled out as the origin for the FIG that was reproducibly observed in our tunnel junctions on $CeCoIn_5$, as discussed in the main text. Here, we provide an additional check on the possibility of surface $Ce^{3+}$ ions acting like Kondo impurities via an analysis of the ZBCP using the Frota function [13]. Figure S13 (a) displays normalized conductance from a nodal junction at 0.4 K. From the best fit to the data, the Kondo temperature ($T_K$) is found to be 10.2 K, much higher than 1.7 K for diluted Ce impurities in $Ce_{1-x}La_xCoIn_5$ [14]. Furthermore, the ZBCP doesn't show the expected temperature dependence for the width of a Kondo resonance in the strong coupling regime in a Fermi liquid [12,15], as shown in Fig. S13 (b). And the $T_K$ values extracted from the best fits are even higher. Thus, taken together with the observation from Fig. S10, the ZBCP must be of more bulk Kondo lattice origin.

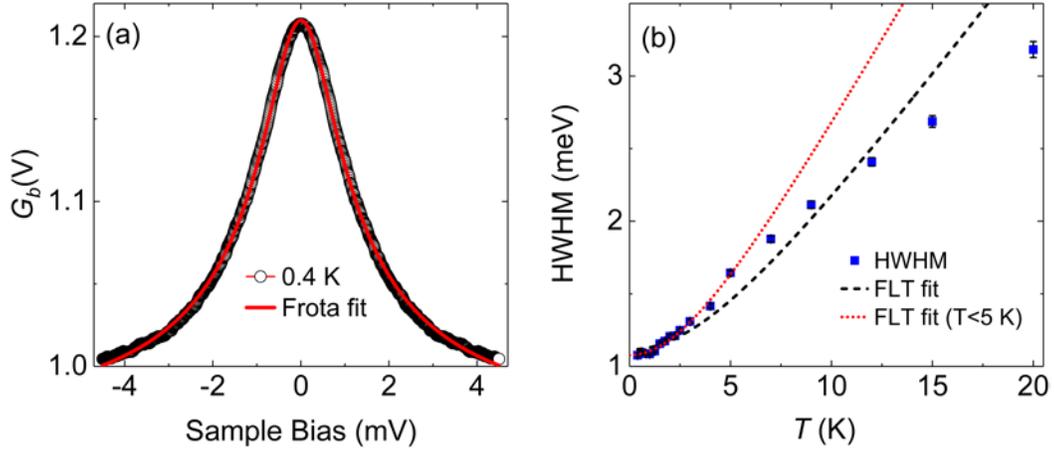

**Figure S13. Checking on the possibility of a single impurity Kondo resonance as the origin of the ZBCP.** (a) Normalized conductance across a (110) junction at 0.4 K. The solid line is the best-fit to the Frota function (Ref. 13): $G_{\text{fit}}(V) = 0.951 + 0.259 \times Re\sqrt{\left(\frac{0.879\times10^{-3}i}{V+0.879\times10^{-3}i}\right)}$. (b) Temperature evolution of the ZBCP's half-width at half-maximum (HWHM) from the junction shown in Fig. S10. The black dashed and red dotted lines are the best fit curves using the Fermi liquid theory (FLT, Refs. 12 & 15) for the half-width at half maximum, $HWHM = \frac{1}{2}\sqrt{(\alpha k_B T)^2 + (2k_B T_K)^2}$, with $T_K = 12.9$ K & $\alpha = 4.3$ and $T_K = 12.5$ K & $\alpha = 5.7$, respectively.

## 8. Pairing gap signature above $H_{cr}$

As shown in the main text, the pairing gap appears to be extrapolated to zero at $H_{c2}$ in both (001) and (100) junctions. It is an interesting/important point whether the pairing gap signature can be actually observed well beyond the crossover region. Figure S14 (a) shows the magnetic field dependence of a junction on the (100) surface at T = 20 mK. As usual, the SC gap is gradually suppressed with the field and the FIG appears above $H_{cr}$ = 4 T. On a close look at the data near $H_{c2}$ (11.8 T) in the low bias region, two slopes are discernible at 9 T as indicated by the dashed lines but the curve exhibits only one slope at 12 T, as shown in Fig. S14 (b). The disappearance of the low-bias kink at 12 T suggests that it is a signature of the pairing gap. We determined the slope change position as a function of field, $V_{ps}(H)$, which was included in Fig. 4(h) in the main text.

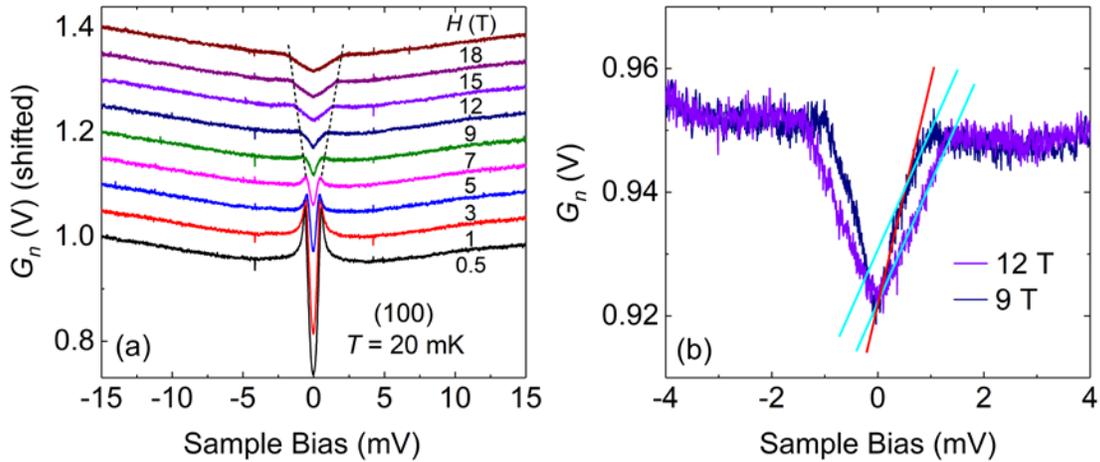

**Figure S14. Field evolution of the SC gap along [100].** (a) Normalized conductance of a (100) junction at $T$ = 20 mK, showing the FIG behavior at higher fields. Due to the dominant effect of the FIG above 4 T, the pairing gap feature is not clearly observed in the data. (b) The conductance curves near $H_{c2}$ (= 11.8 T) at low bias. Within the FIG, there exist two slopes at 9 T (below $H_{c2}$), which turns into a single slope at 12 T (above $H_{c2}$), as indicated by the solid lines. This slope change within the FIG is due to the pairing gap.

## 9. Estimation of the Landé g-factor

As shown in the main text, the linear field dependence of the FIG is reminiscent of the Zeeman effect. To estimate the Landé g-factor, we used $V_s$ instead of $V_p$ as adopted in the literature [10,12]. This can also justified by simulation shown in Fig. S15(a). Figure S15(b) illustrates how we determined $V_s$, that is, by averaging the two bias voltages for the maximum and minimum in $dG_b/dV$. The error bar in $V_s$ was determined manually by inspecting the shape of the two extrema.

## 10. Robustness of the FIG

As seen in Fig. 2 in the main text, the FIG in CeCoIn$_5$ is reproducibly observed in all three major crystallographic directions. Here, we provide further evidence for its robustness. First, the FIG appears regardless of the junction location on the polished crystal surface, as shown in Fig. S16. All three junctions show the SC pairing gap at low fields, gradually turning into the FIG at high fields. Second, the FIG is not only observed in all CeCoIn$_5$ crystals grown in the same batch but also by three independent groups, as shown in Fig. S17. Taken all together, the FIG must be related to the intrinsic superconductivity in CeCoIn$_5$.

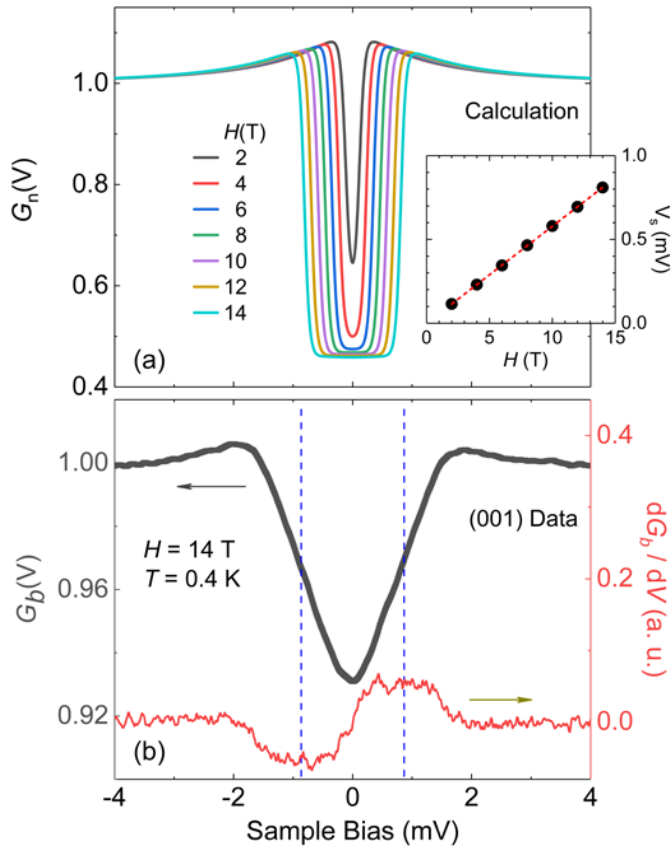

**Figure S15. Estimation of the Landé g-factor.** (a) Calculated tunneling conductance at $T = 0.4$ K based on the AA theory using the expressions for $G_2$ and $G_3$ terms in Ref. 10 with the weight factor of 0.5 per each. The g-value is set to 2 and the spin is set to 1/2. Inset: $V_s$ determined from the steepest point in each curve. The red dashed line is a linear fit, from which g-factor is estimated to be 2.0055±0.0048, very close to the input value. This shows that taking Vs is a more accurate way to determine the g-factor that taking the nominal peak position. (b) Normalized conductance for a (001) junction at $T = 0.4$ K and $H = 14$ T and its first derivative (red line, right axis). The maximum and the minimum in the first derivative correspond to points where the FIG shows the steepest slope, as indicated by the dashed lines.

Additional evidence for its intrinsic nature can also be found in the literature: e.g., a gap-like feature observed above $H_{c2}$ in STS measurements [16,17].

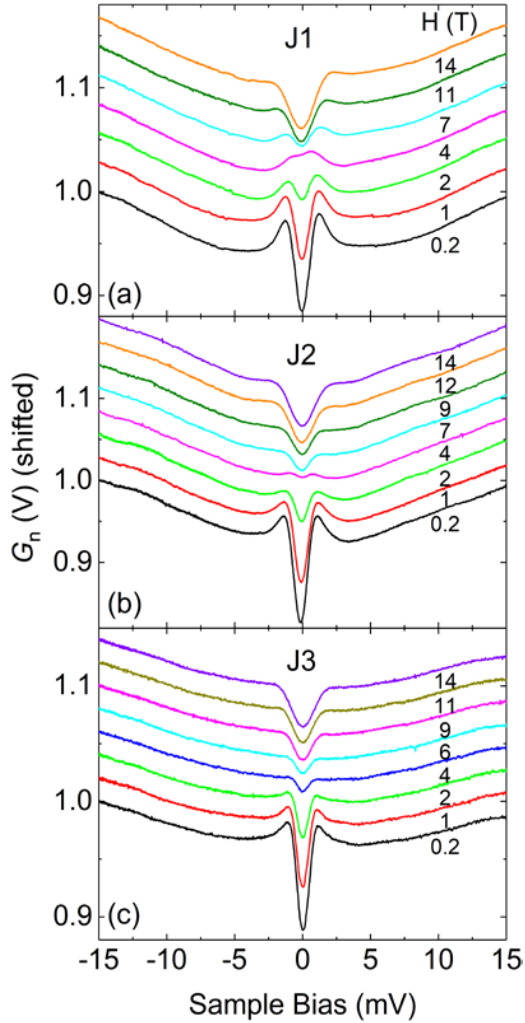 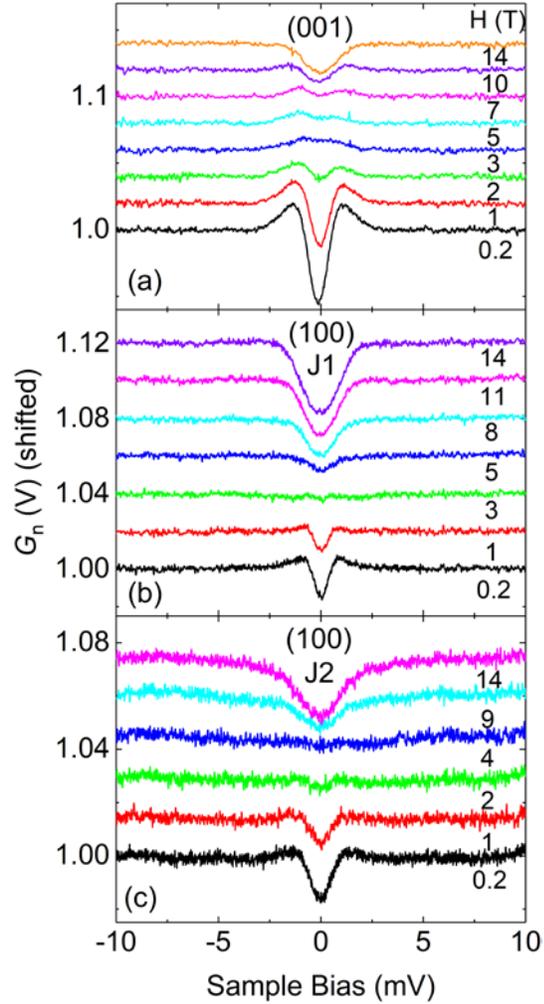

**Figure S16. Robustness of the FIG.** Comparison of normalized conductance spectra at 1.75 K from three (001) tunnel junctions (J1, J2, and J3) prepared at three different locations (roughly 400 μm apart from each other) on the same single crystal. The smooth evolution of the SC gap into the FIG reported in the main text is commonly observed in all three junctions, as shown in (a), (b) and (c).

**Figure S17. Sample independence of the FIG.** Comparison of normalized conductance spectra at 1.75 K from tunnel junctions prepared on three single crystals grown by different groups. The smooth evolution of the SC gap into the FIG reported in the main text is commonly observed, as shown in (a), (b), and (c). The variation in the sharpness is not due to the crystal quality but junction-specific processing conditions.

## 11. Tunneling signature characteristic of spin-fluctuation mediated pairing

According to the Bardeen-Cooper-Schrieffer theory of superconductivity, two electrons pair up to form a Cooper pair due to the mediation by phonons. One of the decisive evidences for this pairing mechanism was found from PTS on Pb, in which the conductance shows two characteristic hump-dip structures outside coherence peaks at a bias voltage of $\Delta + \Omega_{ph}$ (or at $-\Delta - \Omega_{ph}$), where $\Delta$ and $\Omega_{ph}$ represent the SC gap and the energy of the phonon mode, respectively [18-21]. From the analysis of these structures along with more detailed second derivative spectra ($d^2I/dV^2$), it was unambiguously determined that two acoustic phonon modes are involved in the pairing. In many unconventional superconductors, spin fluctuations are widely believed to play a similar role to phonons in the Cooper pairing process. There has been ample evidence reported in the literature including the ubiquitously observed neutron spin resonance. Also, the corresponding signature has been frequently observed in tunneling conductance, namely, a hump-dip structure at $\Delta + \Omega_{res}$ outside the SC gap. Apparently, this hump-dip structure is similar to what was observed in Pb as mentioned above, suggesting that Cooper pairs in these superconductors are formed via exchange of some sort of collective excitations of energy $\sim \Omega_{res}$, that is, spin fluctuations in line with the observation of spin resonance [22-26].

## 12. Field-angle dependence of the FIG

The FIG doesn't exhibit any noticeable field angle dependence, as shown in Fig. S18 in the high-field limit. The $\chi$ in CeCoIn$_5$ is anisotropic in its magnitude but its behavior is qualitatively similar for H // c and H // ab in that it persists to follow the Curie-Weiss temperature dependence even below $T_{coh}$ [8], unlike many other heavy fermions, and that it undergoes slope decrease below ~$T_p$ in the $1/T$ plot (Fig. S7). As discussed throughout this paper, simply speaking, the FIG originates from Zeeman splitting of the Ce $4f^1$ moments which are the major contributors to $\chi$. Thus, the absence of field angle dependence of the FIG supports that the FIG is an intrinsic property of CeCoIn$_5$.

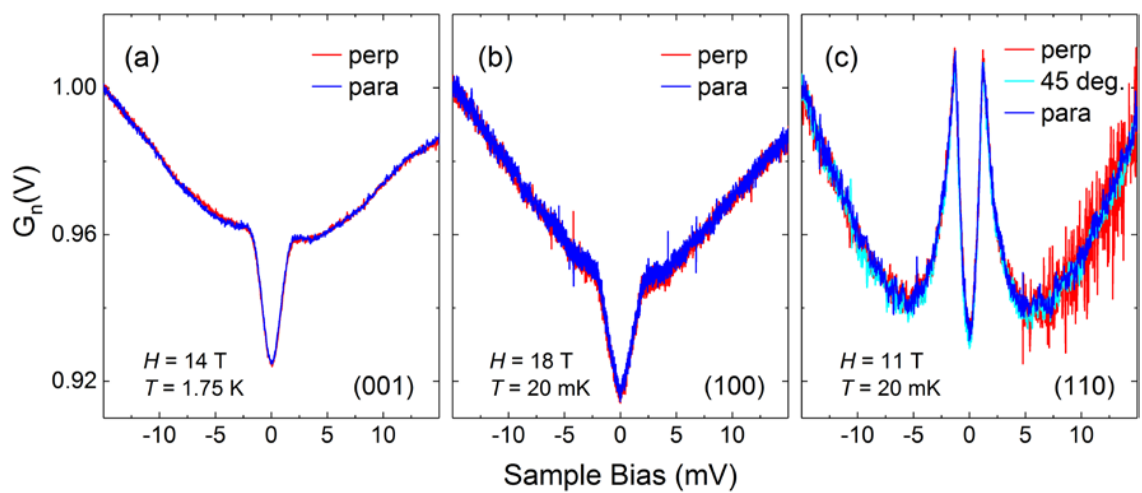

**Figure S18. Field angle dependence of the FIG in CeCoIn$_5$.** (a) (001) junction, (b) (100) junction, and (c) (110) junction. In the legends, 'perp' ('para') indicates that the magnetic field is applied perpendicular (parallel) to the junction plane. No noticeable dependence of the FIG on the field angle is seen in all three junctions.